\documentclass[aps,prb,amsmath,amssymb,showpacs,twocolumn,superscriptaddress]{revtex4-2}
\usepackage{amsmath}
\usepackage{amssymb}
\usepackage{amsthm}
\usepackage{setspace}
\usepackage{graphicx}
\usepackage{braket}
\usepackage{mathrsfs}
\usepackage{float}
\usepackage[colorlinks = true,linkcolor = blue,urlcolor  = blue,citecolor = blue,anchorcolor = blue]{hyperref}
\usepackage[utf8]{inputenc}
\usepackage[english]{babel}
\usepackage{bm}
\usepackage{csquotes}

\begin{document}

\title{Strain-induced Ettingshausen effect in spin-orbit coupled noncentrosymmetric metals }
\author{Gautham Varma K.}
\affiliation{School of Physical Sciences, Indian Institute of Technology Mandi, Mandi 175005, India.}
\author{Azaz Ahmad}
\affiliation{School of Physical Sciences, Indian Institute of Technology Mandi, Mandi 175005, India.}
\author{Gargee Sharma}
\affiliation{Department of Physics, Indian Institute of Technology Delhi, Hauz Khas, New Delhi 110016, India.}

\begin{abstract}
Elastic deformations couple with electronic degrees of freedom in materials to generate gauge fields that lead to interesting transport properties. Recently, it has been well studied that strain-induced chiral magnetic fields in Weyl semimetals lead to interesting magnetotransport induced by the chiral anomaly (CA). Recent studies have revealed that 
CA is not necessarily only a Weyl-node property, but is rather a Fermi surface property, and is also 
present in a more general class of materials, for example, in spin-orbit-coupled noncentrosymmetric metals (SOC-NCMs). The interplay of strain, CA, and charge and thermomagnetic transport in SOC-NCMs, however, remains unexplored. Here we resolve this gap. Using a tight-binding model for SOC-NCMs, we first demonstrate that strain in SOC-NCMs induces anisotropy in the spin-orbit coupling and generates an axial electric field. Then, using the quasi-classical Boltzmann transport formalism with momentum-dependent intraband and interband scattering processes, we show that strain in the presence of external magnetic field can generate temperature gradients via the Nernst-Ettingshausen effect, whose direction and behavior depends the on interplay of multiple factors: the angle between the applied strain and magnetic field, the presence of the chiral anomaly, the Lorentz force, and the strength of interband scattering. We further reveal that time-reversal symmetry breaking in the presence of an external magnetic field generates the Berry-curvature-driven anomalous Ettingshausen effect, which is qualitatively distinct from the conventional Lorentz-force-driven counterpart. In light of recent and forthcoming theoretical and experimental advances in the field of SOC-NCMs, we find our study to be particularly timely and relevant.
\end{abstract}

\maketitle

\section{Introduction}
Electrons and atomic lattices are the essential building blocks of condensed matter. Fundamentally, the two are different from each other, but their coupling gives rise to a wide range of emergent phenomena. 
In the last decade, a large body of work has shown that elastic deformations (strain) in certain lattices couple to the electronic degrees as an axial or chiral magnetic field. The occurrence of such fields has been extensively studied in graphene~\cite{manes2007symmetry,manes2013generalized,iorio2015revisiting,sohier2014phonon,vozmediano2010gauge,castro2009electronic}, and their presence has been verified in STM measurements~\cite{low2010strain,guineaenergy,settnes2017valley}. Similarly, chiral magnetic fields were also shown to emerge in three-dimensional topological Dirac and Weyl semimetals~\cite{cortijo2015elastic,cortijo2016strain,shapourian2015viscoelastic}. The occurrence of such gauge fields in Weyl semimetals is of profound importance due to the implications on their topological properties, such as the anomalous Hall~\cite{burkov2014anomalous} and Nernst effects~\cite{sharma2016nernst,sharma2017nernst}, open Fermi
arcs~\cite{wan2011topological}, planar Hall and Nernst effects~\cite{nandy2017chiral}, and the manifestation of the much celebrated chiral anomaly (CA), also known as the Adler-Bell-Jackiw anomaly~\cite{adler1969axial,aji2012adler,nielsen1981no,nielsen1983adler,bell1969pcac,goswami2013axionic,goswami2015optical,goswami2013chiral,fukushima2008chiral,ahmad2021longitudinal,ahmad2023longitudinal,ahmad2024geometry,ahmad2025longitudinal,Varma2025chiral}. 
Chiral anomaly refers to the breakdown of chiral charge conservation in the presence of externally applied gauge fields. This manifests itself in transport properties such as longitudinal magnetoconductance (LMC) and planar Hall conductance (PHC). In recent years, there have been investigations on how the anomaly and the associated transport properties vary if the external fields are induced by strain~\cite{grushin2016inhomogeneous,ghosh2020chirality,ahmad2023longitudinal}.

Recently, CA has been proposed for a larger class of materials known as spin-orbit coupled noncentrosymmetric metals (SOC-NCMs) \cite{cheon2022chiral}. SOC-NCMs are chiral crystals with broken inversion symmetries. Their low-energy Hamiltonian contains a $\mathbf{k} \cdot \boldsymbol{\sigma}$ term similar to that in Weyl semimetals (WSMs); however, unlike in WSMs, the $\boldsymbol{\sigma}$ matrices here correspond to the real spin degrees of freedom. Another notable feature of these systems is that they have two Fermi surfaces (FSs) associated with a single band touching point. These two FSs originating from the two bands meeting at a nodal point have equal and opposite non-zero Berry curvature fluxes piercing through them. Thus, interband scattering leads to the notion of chiral anomaly in SOC-NCMs, indicating that, in solids, chiral anomaly may be better understood as a property of the Fermi surface rather than of the node itself. Such understanding has been instrumental in driving subsequent advances in magnetotransport studies of these systems~\cite{varma2024magnetotransport,ahmad2025chiral,das2023chiral,gopalakrishnan2025chiral}. For instance, SOC-NCMs exhibit a universally positive longitudinal magnetoconductance, irrespective of interband scattering, stemming from identical orbital magnetic moments on both Fermi surfaces, unlike the opposite moments in Weyl semimetals~\cite{varma2024magnetotransport}. Furthermore, their anomalous Hall conductivity shows a Zeeman-driven, non-monotonic magnetic-field dependence \cite{varma2024magnetotransport}. How strain affects the corresponding topological properties in SOC-NCMs remains an outstanding question of interest that has received little attention.

In this work, using an exact tight-binding model of a SOC-NCM, we show that strain has a twofold effect: (i) it induces anisotropy in the spin-orbit coupling, thereby mixing the well-defined chiralities of the unperturbed bands, and (ii) an inhomogeneous strain generates a chiral electric field $\mathbf{E}_5$. We solve the Boltzmann transport equation for the anisotropic system semi-analytically, augmented with charge conservation and momentum-dependent interband and intraband scattering processes, and show that strain-induced LMC behaves quadratically with the external magnetic field and behaves linearly with the applied strain. 

We next turn to the Nernst–Ettingshausen effect. In the Ettingshausen effect, a transverse temperature gradient is generated as a result of an electrical current in the presence of an external magnetic field. This is the converse of the Nernst effect, where a temperature gradient induces a transverse voltage~\cite{ri1994nernst,behnia2016nernst,behnia2007nernst}. The Nernst effect has been extensively studied in a wide range of materials, from high-$T_c$ superconductors~\cite{cyr2009enhancement,tinh2009theory,wang2006nernst} to Dirac and Weyl semimetals~\cite{das2023chiral,sharma2016nernst,sharma2017nernst}. More recently, the anomalous Nernst effect has been observed in time-reversal-symmetry-broken Weyl semimetals~\cite{liang2017anomalous,zhou2022anomalous,roychowdhury2023anomalous}. In this work, we investigate the strain-induced Ettingshausen effect in SOC-NCMs, revealing the nature and directionality of the resulting thermal gradients arising from the interplay of multiple factors: the angle between the applied strain and magnetic field, the presence of the chiral anomaly, the Lorentz force, and the strength of interband scattering. We further identify the presence of the anomalous Ettingshausen effect, driven by Berry curvature when time-reversal symmetry is broken in the presence of an external magnetic field. With recent and imminent advances in both theory and experiment on SOC-NCMs, our study is both timely and relevant.

Our paper is organized as follows. In Sec.~\ref{2a}-\ref{2c}, we introduce a lattice model of a generic SOC-NCM which is discretized on a simple cubic lattice, and incorporate the effect of external strain. In Sec.~\ref{2d}, we discuss the quasi-classical Boltzmann transport formalism incorporating momentum-dependent scattering at finite temperatures to calculate the charge and thermoelectric coefficients. In Sec.~\ref{3}, we present our results. We study strain-induced LMC, and the behavior of temperature gradients for three different cases: (i) when $\mathbf{E_5 \parallel B}$, (ii) when $\mathbf{E_5 \perp B}$, and (iii) for arbitrary angles. We also discuss the effect of anomalous Hall and Peltier coefficients on the nature of temperature gradients. We conclude our discussions in section Sec.~\ref{4}.

\section{Model and formalism} \label{2}
\subsection{Lattice model of a noncentrosymmetric spin-orbit coupled semimetal} \label{2a}
A lattice model of a generic spin-orbit coupled noncentrosymmetric metal discretized on a simple cubic lattice is written as~\cite{mukherjee2012order}: 
\begin{align}
H=\sum_{\mathbf{k}}\psi_{\mathbf{k}}^{\dagger}[n_{0,\mathbf{k}}\sigma_0 + \mathbf{n}_\mathbf{k}\cdot\boldsymbol{\sigma}]\psi_{\mathbf{k}},
\label{Hsocncm}
\end{align}
where $\psi^\dagger_\mathbf{k} = [c^\dagger_{\uparrow\mathbf{k}}, c^\dagger_{\downarrow\mathbf{k}}]$, and the second-quantized operator $c^\dagger_{s\mathbf{k}}$ create an electron with spin $s$ and momentum $\hbar\mathbf{k}$; $\boldsymbol{\sigma}$ is the vector of Pauli matrices representing spin, and $\sigma_0$ is an identity matrix.
The functions $n_{0,\mathbf{k}}$ and $\mathbf{n}_\mathbf{k}$ are:
\begin{align}
n_{0,\mathbf{k}}& = 2t_1 [\cos(ak_x) + \cos(ak_y) + \cos(ak_z)] \notag \\
    & + 4t_2 [\cos(ak_x) \cos(ak_y) + \cos(ak_y) \cos(ak_z) 
    \notag\\ &\hspace{0.7cm}+ \cos(ak_z) \cos(ak_x)] \notag \\
    & + 8t_3 \cos(ak_x) \cos(ak_y) \cos(ak_z), \notag \\
\mathbf{n}_{\mathbf{k}}&=t_\mathbf{so}[\sin{(a k_{x}), \sin(ak_y), \sin(ak_z)]}.
\end{align}
In the above $t_1$, $t_2$, and $t_3$ are the spin-independent hopping integrals to the first, second, and third nearest neighbors; $t_\mathbf{so}$ is the spin-orbit coupling strength that enters as a spin-dependent hopping in the tight-binding description. The energy dispersion is evaluated to be:
\begin{align}
    \epsilon_\mathbf{k} = n_{0,\mathbf{k}}\pm |\mathbf{n}_\mathbf{k}|.
\end{align}
Clearly the two bands touch each other at the eight high-symmetry points: $\Gamma:(0,0,0)$, $M:(\pi,\pi,0)$, $(\pi,0,\pi)$, $(0,\pi,\pi)$, $X:(\pi,0,0)$, $(0,\pi,0)$, $(0,0,\pi)$, $R:(\pi,\pi,\pi)$. Around these points, the Hamiltonian  can be expanded in the following form (see Appendix A):
\begin{align}
    h_\mathbf{k} = \epsilon_{0\mathbf{k}} \pm at_{\text{so}} \mathbf{k} \cdot \boldsymbol{\sigma}.
\end{align}
The low-energy Hamiltonian around the eight symmetry points resembles a Weyl fermion ($\sim \mathbf{k}\cdot\boldsymbol{\sigma}$). Out of these eight points, four of them ($\Gamma$ and $M$ points) are of positive chirality, and the other four points ($X$ and $R$ points) are of negative chirality. The $X$ points occur at energy: $\epsilon_{X}=-4t_2 - 8t_3 + 2 t$, and all the $M$ points occur at energy: $\epsilon_{M}=-4t_2 + 8 t_3 - 2t$. The $\Gamma$ and $R$ points occur at energy $\epsilon_\Gamma = 6t_1+12t_2+8t_3$, and $\epsilon_R = -6t_1-4t_2-8t_3$, respectively. 
The energy difference between the positive and negative chirality points is $\Delta\epsilon = 64t_3$. Therefore, for typical values of the tightbinding parameters relevant to noncentrosymmetric SOC metals, it is sufficient to consider the low-energy Hamiltonian only around one band-touching point of a particular chirality~\cite{mukherjee2012order,cheon2022chiral}.  

\subsection{Effect of strain} \label{2b}
We now discuss the application of strain to the above model following Ref.~\cite{vozmediano2010gauge,shapourian2015viscoelastic,cortijo2015elastic,cortijo2016strain,castro2009electronic}. When an inhomogeneous strain $\mathbf{U}(\mathbf{r})$ is applied to the lattice, the hopping element $t_{\mathbf{R}+\boldsymbol{\delta}_{i}, \mathbf{R}}$ between sites $\mathbf{R}$ and $\mathbf{R}+\boldsymbol{\delta}_i$ is modified as a
result of the lattice distortion, i.e., $t_{\mathbf{R}+\boldsymbol{\delta}_{i}, \mathbf{R}}\rightarrow t_{\mathbf{R}+\boldsymbol{\delta}_{i}, \mathbf{R}}+ \delta t_{\mathbf{R}+\boldsymbol{\delta}_{i}, \mathbf{R}}$, where the change in the hopping parameter $\delta t_{\mathbf{R}+\boldsymbol{\delta}_{i}, \mathbf{R}}$ is
\begin{align}
    \delta t_{\mathbf{R}+\boldsymbol{\delta}_{i}, \mathbf{R}} \approx 
    -\sum_{\mu\nu} \delta_i^{\mu} u_{\mu\nu} \delta_i^\nu \frac{t_{\mathbf{R}+\boldsymbol{\delta}_{i}, \mathbf{R}}} {|\boldsymbol{\delta}_{i}|^2}\left(-\frac{\partial \ln t(\mathbf{r})}{\partial \mathbf{r}}\right),
\end{align}
where the summation is over all spatial indices, and $u_{\mu\nu}$ are the elements of the strain tensor: $u_{\mu\nu} = (\partial_\mu U_\nu+\partial_\nu U_\mu)/2$. We first consider a uniaxial homogeneous strain along the $z$-axis: $\mathbf{U} = (0,0, dz/\beta)$, and therefore $u_{zz} = d/\beta$, and all other components of the strain tensor vanish ($\beta = (-{\partial \ln t(\mathbf{r})}/{\partial \mathbf{r}}$ being the Gr\"uneisen parameter). Incorporating changes in the hopping parameters, the strained Hamiltonian now becomes
\begin{align}
H'=\sum_{\mathbf{k}}\psi_{\mathbf{k}}^{\dagger}[n'_{0,\mathbf{k}}\sigma_0 + \mathbf{n}'_\mathbf{k}\cdot\boldsymbol{\sigma}]\psi_{\mathbf{k}},
\label{Hsocncmstrain}
\end{align}
where 
\begin{align}
n'_{0,\mathbf{k}}& = 2[t_1\cos(ak_x) + t_1\cos(ak_y) + t'_1\cos(ak_z)] \notag \\
    & + 4[t_2\cos(ak_x) \cos(ak_y) + t'_2\cos(ak_y) \cos(ak_z) 
    \notag\\ &\hspace{0.7cm}+ t'_2\cos(ak_z) \cos(ak_x)] \notag \\
    & + 8t'_3 \cos(ak_x) \cos(ak_y) \cos(ak_z),\nonumber\\
\mathbf{n}'_{\mathbf{k}}&=[t_\mathbf{so}\sin{(a k_{x}),t_\mathbf{so}\sin(ak_y), t'_\mathbf{so}\sin(ak_z)]},
\end{align}
{where the modified hoppings are:}
\begin{align}
t'_1 &= t_1(1-d/a^2),\quad t'_2 = t_2(1-(d)(\gamma/2a^2\beta)),\nonumber\\
t'_3 &=t_3 (1-(d)(\omega/3a^2\beta)), t'_\mathbf{so}=t_\mathbf{so}(1-d/a^2),\nonumber\\ 
\beta &= -\partial \ln t_1/\partial r, \quad 
\gamma = -\partial \ln t_2/\partial r, \quad \omega = -\partial \ln t_3/\partial r.
\end{align}
Since SOC-NCMs have only one relevant band touching point near the Fermi surface, we expand the strained low-energy Hamiltonian around the $\Gamma$ point as 
\begin{align}
   h^\Gamma_\mathbf{k} = \epsilon^\Gamma_\mathbf{k}\sigma_0  + at_\mathbf{so}\mathbf{k}\cdot\boldsymbol{\sigma} + b_z{k_z}\sigma_z +b_0\sigma_0,
\end{align}
with $b_z$ and $b_0$ being the strain-dependent parts that arise due to the modified hopping parameters. 
The effect of strain is therefore twofold: a uniform energy shift, represented by the \( b_0\sigma_0 \) term, and an anisotropy in the spin--orbit coupling arising from the \( b_z\sigma_z \) term.

We further note that the strain-induced axial magnetic field $\mathbf{B}_5$, which depends on the curl of the separation between positive and negative chirality nodes, vanishes in this model as the respective positions of the band-touching points in the Brillouin zone remain unaffected by strain.
However, an inhomogeneous or a time-dependent strain profile (that may be generated by dynamically stretching and
compressing the sample) can generate an axial electric field term $\mathbf{E}_5=-\partial_t b_0-\partial_z b_z$~\cite{cortijo2015elastic,cortijo2016strain,grushin2016inhomogeneous}. 
\subsection{Effective Hamiltonian} \label{2c}
It suffices to consider the following low-energy effective Hamiltonian for our remaining analysis: 
\begin{align}
    h^\Gamma_\mathbf{k} = \frac{\hbar^2}{2m} k^2 + a t_\mathbf{so} \mathbf{k}\cdot\boldsymbol{\sigma} + b_z{k_z}\sigma_z.
    \label{Eq:heffgamma}
\end{align}
The constant energy shift can be ignored, and the effect of inhomogeneous or dynamic strain is encoded in an axial electric field $\mathbf{E}_5 = (0,0,E_5)$. Since the $b_z$ term is relatively small in Eq.~\ref{Eq:heffgamma}, its effect is included perturbatively up to first order. Such a perturbation expansion also allows us to proceed analytically, as the calculation otherwise becomes intractable. 
The energy dispersion of the effective Hamiltonian is the following: 
\begin{align}
    \epsilon_\mathbf{k}^{\pm} = \frac{\hbar^2k^2}{2m} \pm at_\mathbf{so} k\pm b_z k \cos^2\theta, 
\end{align}
where $\theta$ is the polar angle. The eigenstates of the effective Hamiltonian are given by: 
\begin{align}
     |u_\mathbf{k}^+\rangle &= 
\begin{pmatrix}
e^{-i\phi} \cos\theta / 2 \\
\sin\theta / 2
\end{pmatrix}
- b'_z\sin 2\theta 
\begin{pmatrix}
-e^{-i\phi} \sin\theta / 2 \\
\cos\theta / 2
\end{pmatrix}
\nonumber\\
|u_\mathbf{k}^-\rangle &= 
\begin{pmatrix}
-e^{-i\phi} \sin\theta / 2 \\
\cos\theta / 2
\end{pmatrix}
+ b'_z\sin 2\theta 
\begin{pmatrix}
e^{-i\phi} \cos\theta / 2 \\
\sin\theta / 2
\end{pmatrix},
\end{align}
where $\phi$ is the azimuthal angle, and $b'_z={b_z}/{4at_\mathbf{so}}$. In the absence of strain, the eigenstates of the Hamiltonian are identical to those of a chiral Weyl fermion ($\mathbf{k}\cdot\boldsymbol{\sigma}$). Strain, which effectively makes the Weyl fermion anisotropic, therefore, mixes the well-defined chirality of the unperturbed bands. The Berry curvature ($\boldsymbol{\Omega}_\mathbf{k}$) and the orbital magnetic moment ($m_\mathbf{k}$) of the Bloch bands of a Hamiltonian $H_\mathbf{k}$ can be evaluated by the following expressions~\cite{xiao2010berry,hagedorn1980semiclassical,chang1996berry}: $\boldsymbol{\Omega}_\mathbf{k} = i \nabla_{\mathbf{k}} \times \langle u_\mathbf{k} | \nabla_{\mathbf{k}} | u_\mathbf{k} \rangle$, and $\mathbf{m}_\mathbf{k}= -({ie}/{2\hbar}) \text{Im} \langle \nabla_{\mathbf{k}} u_\mathbf{k}|[ \epsilon(\mathbf{k}) - {H_\mathbf{k}} ]| \nabla_{\mathbf{k}} u_\mathbf{k}\rangle$.
For our effective Hamiltonian, they are evaluated up to first order in $b_z^{'}$ to be:
\begin{align}
\Omega_\mathbf{k}^+ &= \frac{2b_z' + 6 b_z' \cos(2\theta)-1}{{2k^2}}\hat{\mathbf{k}} \nonumber\\
\Omega_\mathbf{k}^- &= \frac{-2b_z' - 6 b_z' \cos(2\theta)+1}{{2k^2}}\hat{\mathbf{k}} \nonumber\\ 
       m^+_\mathbf{k} &=\frac{e a t_\mathrm{so}}{2\hbar k} (1 - 4b_z'\cos 2\theta) \hat{\mathbf{k}}\nonumber\\
    m^-_\mathbf{k}&=\frac{e a t_\mathrm{so}}{2\hbar k} (1 - 4b_z'\cos 2\theta) \hat{\mathbf{k}}
   \end{align}
\\
It is straightforward to verify that in absence of the anisotropic term ($b'_z$), the above expressions reduce to the well-known expressions: $\boldsymbol{\Omega}^\lambda_\mathbf{k}=-\lambda \mathbf{k} /2k^3$, and $\mathbf{m}^{\lambda}_\mathbf{k}=e a t_\mathbf{so} \mathbf{k}/{2\hbar k^2}$. Note that the Berry curvature changes sign for both bands, but the orbital magnetic moment does not! This feature, which is true in the absence of strain~\cite{varma2024magnetotransport}, also holds in the current case, i.e., in the presence of strain. 
\subsection{Boltzmann transport formalism} \label{2d}
Due to the effect of Berry phase, in the presence of electric and magnetic fields, the semiclassical dynamics of the Bloch electrons are modified and governed by the following set of equations~\cite {son2012berry}:.
\begin{align}
\dot{\mathbf{r}}^\lambda &= \mathcal{D}^\lambda \left( \frac{e}{\hbar}\left(\mathbf{E}\times \boldsymbol{\Omega}^\lambda \right)+ \frac{e}{\hbar}(\mathbf{v}^\lambda\cdot \boldsymbol{\Omega}^\lambda) \mathbf{B} + \mathbf{v}_\mathbf{k}^\lambda\right) \nonumber\\
\dot{\mathbf{p}}^\lambda &= -e \mathcal{D}^\lambda \left( \mathbf{E} + \mathbf{v}_\mathbf{k}^\lambda \times \mathbf{B} + \frac{e}{\hbar} \left(\mathbf{E}\cdot\mathbf{B}\right) \boldsymbol{\Omega}^\lambda \right)
\label{Couplled_equation}
\end{align}
where $\mathbf{v}_\mathbf{k}^\lambda = \frac{1}{\hbar}\frac{\partial\epsilon^{\lambda}(\mathbf{k})}{\partial\mathbf{k}}$ is the band velocity, $\boldsymbol{\Omega}^\lambda = -\lambda \mathbf{k} /2k^3$ is the Berry curvature, and $\mathcal{D}^\lambda = (1+e\mathbf{B}\cdot\boldsymbol{\Omega}^\lambda/\hbar)^{-1}$ is the factor modifying the density of the states in the presence of the Berry curvature, and $\lambda$ denotes the band index. The self-rotation of the Bloch wave packet also gives rise to an orbital magnetic moment, $\mathbf{m}^\lambda_\mathbf{k}$~\cite{xiao2010berry}. In the presence of a magnetic field, the orbital magnetic moment shifts the energy dispersion as $\epsilon^{\lambda}_{\mathbf{k}}\rightarrow \epsilon^{\lambda}_{\mathbf{k}} - \mathbf{m}^\lambda_\mathbf{k}\cdot \mathbf{B}$. The Boltzmann transport equation in the steady state is given by \\
\begin{align}
 \dot{\mathbf{r}} \cdot \nabla_{\mathbf{r}} f_{\mathbf{k}}^{\lambda} + \dot{\mathbf{k}} \cdot \nabla_{\mathbf{\mathbf{k}}} f_{\mathbf{k}}^{\lambda} = \mathcal{I}_{\text{coll}},
 \label{Eq_boltz_eqn}
\end{align}
where
\begin{align}\mathcal{I}_{\text{coll}}=\sum\limits_{\lambda'}\sum\limits_{\mathbf{k}'} W^{\lambda\lambda'}_{\mathbf{k}\mathbf{k}'} (g^\lambda_{\mathbf{k}'} - g^\lambda_\mathbf{k}) 
\end{align}
is the collision integral, the Boltzmann distribution function $f^{\chi}_\mathbf{k} = f_{0} + g^{\chi}_{\mathbf{k}}$, where $f_{0}$ is the Fermi-Dirac distribution, and $g^{\chi}_{\mathbf{k}}$ is the deviation. The coefficient $W^{\lambda\lambda'}_{\mathbf{k}\mathbf{k^{'}}}$ is calculated using the Fermi-golden rule. 
\begin{align}
{W}^{\lambda \lambda'}_{\mathbf{k k'}} = \frac{2\pi n}{\hbar\mathcal{V}}|\bra{u^{\lambda'}(\mathbf{k'})}U^{\lambda \lambda'}_{\mathbf{k k'}}\ket{u^{\lambda}(\mathbf{k})}|^2\times\delta(\epsilon^{\lambda'}(\mathbf{k'})-\epsilon^\lambda_\mathbf{k}).\nonumber\\
\label{Eq_Fermi_golden_rule}
\end{align}
Here $n$ is the impurity concentration, $\mathcal{V}$ is the system's volume, and the operator $U^{\lambda \lambda'}_{\mathbf{k k'}}$ describes the potential profile of the scattering centers. 
We assume elastic impurities, and set $U^{\lambda \lambda'}_{\mathbf{k k'}}= I_{3\times3}U^{\lambda \lambda'}$, where the matrix elements $U^{\lambda \lambda'}$ control the interband scattering ($\lambda\neq\lambda'$) and intraband scattering ($\lambda=\lambda'$) independently. The quantity ${W}^{\lambda \lambda'}_{\mathbf{k k'}}$ crucially depends on the overlaps of the Bloch wave-functions that are given by:
\begin{align}
\mathcal{G}^{\lambda \lambda'}&=\frac{1}{2}(1 + \lambda \lambda'(\cos\theta \cos\theta' + \sin\theta \sin\theta' \cos(\phi- \phi') \nonumber\\
+~ &b'_{z} ( 2\cos\theta \cos\theta' - \cos\theta \cos^3\theta' - \cos^3\theta \cos\theta' \nonumber\\
- &~\cos^2\theta \sin\theta \sin\theta'\cos (\phi-\phi')\nonumber\\-&\cos^2\theta' \sin\theta \sin\theta'\cos (\phi-\phi')~)))
\label{Overlap_p1p1}
\end{align}
Since at finite temperatures the steady state distribution function $f_{\mathbf{k}}^\lambda$ is temperature dependent, the first term in the left-hand-side of Eq.~\ref{Eq_boltz_eqn} is non-zero as temperature varies spatially, i.e., $T=T(\mathbf{r})$. 
The steady state Boltzmann transport equation at finite temperature up to linear order in $\mathbf{E}$ and $\nabla {T}$ can be derived to be:
\begin{align}
&\left[\left(\frac{\partial f_{\mathbf{k}}^\lambda}{\partial \epsilon^\lambda_\mathbf{k}}\right) \left(e\mathbf{E}+\frac{\epsilon_{\mathbf{k}}^\lambda-\mu}{T}\nabla T\right)\cdot \left(\mathbf{v}^\lambda_\mathbf{k} + \frac{e\mathbf{B}}{\hbar} (\boldsymbol{\Omega}^\lambda\cdot \mathbf{v}^\lambda_\mathbf{k}) \right)\right]\nonumber\\
 &= -\frac{1}{ \mathcal{D}^\lambda}\sum\limits_{\lambda'}\sum\limits_{\mathbf{k}'} W^{\lambda\lambda'}_{\mathbf{k}\mathbf{k}'} (g^\lambda_{\mathbf{k}'} - g^\lambda_\mathbf{k})
 \label{Eq_boltz2}
\end{align}
We fix the direction of the chiral electric field $(\mathbf{E_{5}})$ along increasing $z$-direction, and apply an external magnetic field in the $xz$-plane. Therefore, $\mathbf{E_5} = E(0,0,1)$ and  $\mathbf{B} = B (\cos{\gamma},0,\sin{\gamma})$. We consider $g_k^{\lambda}$ to be of the form:
\begin{align}
    g_k^{\lambda}=\left(\frac{\partial f_{\mathbf{k}}^\lambda}{\partial \epsilon^\lambda_\mathbf{k}}\right) \left(e\mathbf{E}+\frac{(\epsilon_{\mathbf{k}}-\mu)}{T}\nabla T\right) \cdot \mathbf{\Lambda_k^{\lambda}} \label{18}
\end{align}
Here, $\mathbf{\Lambda}$ has three components (${\Lambda_x}$,${\Lambda_{y}}$ and ${\Lambda_z}$). Therefore, Eq.~\ref{Eq_boltz2} reduces to three separate equations:
\begin{align}
&\mathcal{D}^{\lambda}(k)\left[{v^{\lambda,x}_{\mathbf{k}}}+\frac{eB\cos{\gamma}}{\hbar}(\mathbf{v^{\lambda}_k}\cdot\mathbf{\Omega}^{\lambda}_k)\right]
 \nonumber\\  &=\sum_{\lambda' \mathbf{k}'}{{W}^{\lambda \lambda'}_{\mathbf{k k'}}}{(\Lambda^{\lambda'}_{{k',x'}}-\Lambda^{\lambda}_{{k,x}})}. \nonumber \\ 
&\mathcal{D}^{\lambda}(k){v^{\lambda,y}_{\mathbf{k}}}
= \sum_{\lambda' \mathbf{k}'}{{W}^{\lambda \lambda'}_{\mathbf{k k'}}}{(\Lambda^{\lambda'}_{{k',y'}}-\Lambda^{\lambda}_{{k,y}})}.\nonumber \\ \nonumber
&\mathcal{D}^{\lambda}(k)\left[{v^{\lambda,z}_{\mathbf{k}}}+\frac{eB\sin{\gamma}}{\hbar}(\mathbf{v^{\lambda}_k}\cdot\mathbf{\Omega}^{\lambda}_k)\right]
\nonumber\\&= \sum_{\lambda' \mathbf{k}'}{{W}^{\lambda \lambda'}_{\mathbf{k k'}}}{(\Lambda^{\lambda'}_{{k',z'}}-\Lambda^{\lambda}_{{k,z}})}.
\label{boltzman_in_terms_lambda}  
\end{align}

We define the valley scattering time ($\tau^\lambda_\mathbf{k}$) as follows
\begin{align}
\frac{1}{\tau^{\lambda}_{\mathbf{k}}(\theta,\phi)}=\sum_{\lambda'}\mathcal{V}\int\frac{d^3\mathbf{k'}}{(2\pi)^3}(\mathcal{D}^{\lambda'}_{\mathbf{k}'})^{-1}{W}^{\lambda \lambda'}_{\mathbf{k k'}},
\label{Tau_invers}
\end{align}
where we have taken the Berry phase that changes the density of states as: $\sum_{k}\longrightarrow \mathcal{V}\int\frac{d^3\mathbf{k}}{(2\pi)^3}\mathcal{D}^\lambda(k)$. Eq.~\ref{boltzman_in_terms_lambda} is then rewritten as: 
\begin{multline}
\mathbf{h}^{\lambda}_k(\theta,\phi) + \frac{\mathbf{\Lambda}_k^{\lambda}(\theta,\phi)}{\tau^{\lambda}_k(\theta,\phi)}\\=\sum_{\lambda'}\mathcal{V}\int\frac{d^3\mathbf{k}'}{(2\pi)^3} \mathcal{D}^{\lambda'}(k'){W}^{\lambda \lambda'}_{\mathbf{k k'}}\mathbf{\Lambda}_{k'}^{\lambda'}(\theta',\phi'),
\label{MB_in_term_Wkk'}
\end{multline}
where $\mathbf{h}^{\lambda}_k 
 (\theta,\phi)=\mathcal{D}^{\lambda}_{\mathbf{k}}[\mathbf{v}^{\lambda}_{\mathbf{k}}+\frac{e\mathbf{B}}{\hbar}(\mathbf{\Omega}^{\lambda}_{k}\cdot \mathbf{v}^{\lambda}_{\mathbf{k}})]$ has three components--$h_x$, $h_y$ and $h_z$.
In the finite temperature limit, we consider a volume in the momentum space which is nearby to the Fermi surface, and thus Eq.~\ref{Tau_invers} and Eq.~\ref{MB_in_term_Wkk'} can be expanded to:
\begin{align}
\frac{1}{\tau^{\lambda}_k(\theta,\phi)} = \nonumber\\ \mathcal{V}\sum_{\lambda'} &\Pi^{\lambda\lambda'}\iiint\frac{(k')^3\sin{\theta'}}{|\mathbf{v}^{\lambda'}_{k'}\cdot{\mathbf{k'}^{\lambda'}}|}dk'd\theta'd\phi' \mathcal{G}^{\lambda\lambda'}(D^{\lambda'}_{\mathbf{k'}})^{-1},
\label{Tau_inv_int_thet_phi}
\end{align}
and
\begin{align}
&\mathbf{h}^{\lambda}_k(\theta,\phi) + \frac{\mathbf{\Lambda}_k^{\lambda}(\theta,\phi)}{\tau^{\lambda}_k(\theta,\phi)}=\nonumber\\
&\mathcal{V}\sum_{\lambda'} \Pi^{\lambda\lambda'}\iiint q^{\lambda'}(k',\theta',\phi')\mathcal{G}^{\lambda\lambda'}\mathbf{\Lambda}^{\lambda'} (k',\theta',\phi') dk'd\theta'd\phi' \label{Eq_boltz_final} 
\end{align}
where $\Pi^{\lambda \lambda'} = n|U^{\lambda\lambda'}|^2 / 4\pi^2 \hbar^2$, and 
\begin{align}
q^{\lambda} (\theta,\phi)=\frac{(k)^3}{|\mathbf{v}^\lambda_{\mathbf{k}}\cdot \mathbf{k}^{\lambda}|} \sin\theta (\mathcal{D}^\eta_{\mathbf{k}})^{-1} \tau^\lambda_k(\theta,\phi).
\end{align}
We have thus arrived at the working form of the Boltzmann equation [Eq.~\ref{Eq_boltz_final}]. The next step is to select an appropriate ansatz for \(\Lambda^\lambda_i(k,\theta,\phi)\) that self-consistently satisfies both sides of the equation.
Using the following ansatz: 
\begin{align}
&\frac{\Lambda^\lambda_i(k,\theta,\phi)}{\tau^\lambda_k(\theta,\phi)} = \big[ \, d^\lambda - h^\lambda_i(k,\theta,\phi) 
+ a_{1}^\lambda \cos\theta + a_{2}^\lambda \cos^3\theta \notag \\
& + b_{1}^\lambda \sin\theta\cos\phi 
+ b_{2}^\lambda \sin\theta\sin\phi \notag \\
& + c_{1}^\lambda \cos^2\theta \sin\theta \cos\phi 
+ c_{2}^\lambda \cos^2\theta \sin\theta \sin\phi 
\, \big] \,, 
\end{align}
Eq.~\ref{Eq_boltz_final} is written in the following form:
\begin{flalign}
d^{\lambda}+a_{1}^\lambda \cos\theta  + a_{2}^\lambda \cos^3\theta + b_{1}^\lambda \sin\theta\cos\phi + \nonumber\\ \nonumber b_{2}^\lambda \sin\theta\sin\phi  + c_{1}^\lambda \cos^2\theta \sin \theta \cos\phi \\\nonumber+ c_{2}^\lambda \cos^2\theta \sin \theta\sin\phi\\ \nonumber
=\sum_{\lambda'}\mathcal{V}\Pi^{\lambda\lambda'}\iiint q^{\lambda'}(k',\theta',\phi')dk'd\theta'd\phi'\nonumber\\\times[d^{\lambda'} - h^{\lambda'}_i(k',\theta',\phi') + a_{1}^{\lambda'} \cos\theta'+ a_{2}^{\lambda'} \cos^3\theta'\nonumber \\ \nonumber+ b_{1}^{\lambda'} \sin\theta'\cos\phi' + \nonumber\\ \nonumber b_{2}^{\lambda'} \sin\theta'\sin\phi'  + c_{1}^{\lambda'} \cos^2\theta'\sin \theta' \cos\phi' \nonumber\\\nonumber +  c_{2}^{\lambda'} \cos^2\theta' \sin \theta'\sin\phi'] \\
\label{Boltzman_final}
\end{flalign}
where, $i=(x,y,z)$. When the aforementioned equation is explicitly put out (for each $i$), it appears as simultaneous equations that must be solved for fourteen variables. The particle number conservation provides another restriction:
\begin{align}
\sum\limits_{\lambda}\sum\limits_{\mathbf{k}} g^\lambda_\mathbf{k} = 0
\label{Eq_sumgk}
\end{align} 
For the 14 unknowns ($d^{\pm 1}, a_1^{\pm 1},a_2^{\pm 1} ,b_1^{\pm 1},b_2^{\pm 1}, c_1^{\pm 1},c_2^{\pm 1}$), Eq.~\ref{Boltzman_final} and Eq.~\ref{Eq_sumgk} are simultaneously solved with Eq.~\ref{Tau_inv_int_thet_phi}. Due to the intricate structure of the equations, integrals with respect to $k'$, $\theta'$, and $\phi'$, the solutions of the simultaneous equations are carried out numerically. 
Having calculated $\mathbf{\Lambda_{k}^{\lambda}}$, we substitute it in Eq.~\ref{18} to obtain the current density:
\begin{align}
    \mathbf{J}=-\frac{e}{\mathcal{V}}\sum_{\lambda,k}\dot{\mathbf{r}}f_\mathbf{k}^{\lambda}.
\end{align}
Using the above expression, we find the Hall and Peltier coefficients of the $\hat\sigma$ and $\hat\alpha$ matrices. 
\section{Results and Discussions} \label{3}
\subsection{Longitudinal magnetoconductance}
\begin{figure}
    \centering
    \includegraphics[width=0.9\columnwidth]{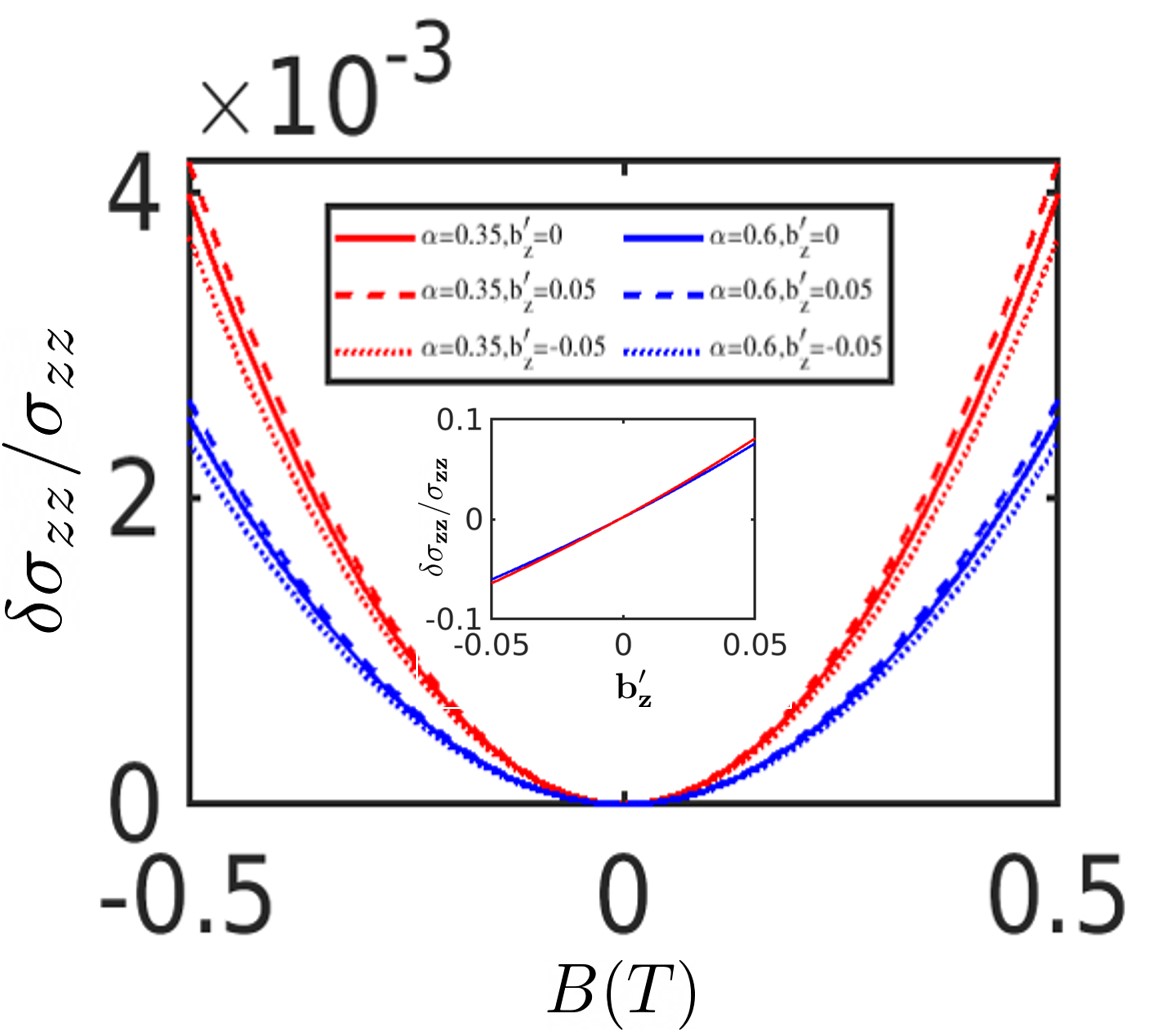}
    \caption {Longitudinal magnetoconductance (LMC) as a function of magnetic field ${B}$ for two different values of interband scattering ($\alpha=0.35$, $\alpha=0.6$) represented by the blue and red curves. The continuous curves denote $b_z^{\prime}=0$. The dashed curves correspond to positive $b_z^{\prime}$, while the dotted ones denote a negative $b_z^{\prime}$. Here, $\delta \sigma_{zz}/\sigma_{zz}=({\sigma_{zz}(B)-\sigma_{zz}(B=0)})/{\sigma_{zz}(B=0)}$. As the inset, we have LMC as a function of the anisotropic coefficient $b_z^{\prime}$ for a fixed value of magnetic field $B=0.5T$, plotted for the interband scattering magnitudes mentioned above. For this case, $\delta \sigma_{zz}/\sigma_{zz}=({\sigma_{zz}(b_z^{\prime})-\sigma_{zz}(b_z^{\prime}=0)})/{\sigma_{zz}(b_z^{\prime}=0)}$.}
\label{Fig:sigmazz}
\end{figure}

In this section, we discuss the main results of this paper. In the previous sections, we noted that a uniaxial strain generates anisotropic spin-orbit coupling in SOC-NCMs, and an inhomogeneous or time-dependent strain can lead to a chiral electric field $E_5$ in the direction of the strain. It is therefore of fundamental interest to first examine the effect of strain on the induced longitudinal magnetoconductance. In the presence of an external magnetic field ($B$), applied in the direction of the external strain, a strain-induced positive longitudinal magnetoconductance is generated, which is quadratic in the magnetic field--as shown in Fig.~\ref{Fig:sigmazz}. Note that this is true even if the anisotropy in the Hamiltonian is ignored, and is generated purely by the $\mathbf{E}_5\cdot\mathbf{B}$ coupling due to the chiral anomaly. 
The strain-induced anisotropy \textit{via} \(b_z^{\prime}\) leads to longitudinal magnetoconductance that remains symmetric about $B=0$, but with a modified quadratic coefficient. This coefficient increases for \(b_z^{\prime} > 0\) and decreases for \(b_z^{\prime} < 0\), reflecting whether the strain compresses or stretches the sample along the $z$-direction. The inset in Fig.~\ref{Fig:sigmazz} shows that LMC varies linearly with strain-induced anisotropy, where the slope slightly varies with interband scattering. If $b_z^{\prime}>0$, compression reduces the lattice parameter, enhancing electron hopping and thereby increasing conductivity. On the other hand, when $b_z^{\prime}<0$, strain increases the lattice parameter, reduces electron mobility, and lowers the conductivity. These results are in stark contrast to the LMC of a strained  Weyl semimetal~\cite{ahmad2023longitudinal}, where the LMC parabola is both tilted and shifted, and the sign of LMC can reverse with increasing intervalley scattering. Such a change of sign in LMC is not observed here as a function of interband scattering (qualitatively same as intervalley scattering), and this critical difference originates due to the same sign of orbital magnetic moment for both the bands in SOC-NCMs, unlike WSMs, where OMM changes sign at both the valleys~\cite{varma2024magnetotransport}. 

\subsection{Ettingshausen effect in SOC-NCMs}
Using a tight-binding lattice model under inhomogeneous strain, we have discussed how SOC-NCMs can generate an induced electric field $\mathbf{E_5}$ in the direction of the strain. We now show that in the presence of an external magnetic field $\mathbf{B}$, strain can lead to a temperature gradient in SOC-NCMs {via} the Ettinghausen effect. Interestingly, we also find that the direction of the temperature gradient is influenced by the angle ($\gamma$) between the direction of the induced electric and magnetic field, and also the interband scattering $\alpha$.

The Ettingshausen effect is a measure of the thermal response to an electric field in the presence of an external magnetic field (the inverse process of the Nernst effect). Using three conductivity tensors, $\hat{\sigma}$, $\hat{\alpha}$ and $\hat{l}$, the following linear response relation can be written between the charge current ($\mathbf{J}$),  thermal current ($\mathbf{Q}$) and the electric field ($\mathbf{E}$) and temperature gradient ($-\nabla T$):
\begin{align}
\begin{bmatrix}
\mathbf{J} \\ 
\mathbf{Q} 
\end{bmatrix} =
\begin{bmatrix}
\hat{\sigma} & \hat{\alpha} \\
\hat{\bar{\alpha}} & \hat{l}
\end{bmatrix}
\begin{bmatrix}
\mathbf{E} \\ 
-\nabla{T}
\end{bmatrix}
\end{align}
Here $\hat{\sigma}$ is the charge conductivity tensor, $\hat{\alpha}$ is the tensor of the Peltier (thermoelectric) coefficients, and in the absence of any charge current, $\hat{l}$ is related to the thermal conductivity tensor $\hat{\kappa}$ as $\hat{\kappa} = \hat{l}-\hat{\bar{\alpha}}\hat{\sigma}^{-1}\hat{\alpha}$. 

\begin{figure}
    \centering
    \includegraphics[width=.98\columnwidth]{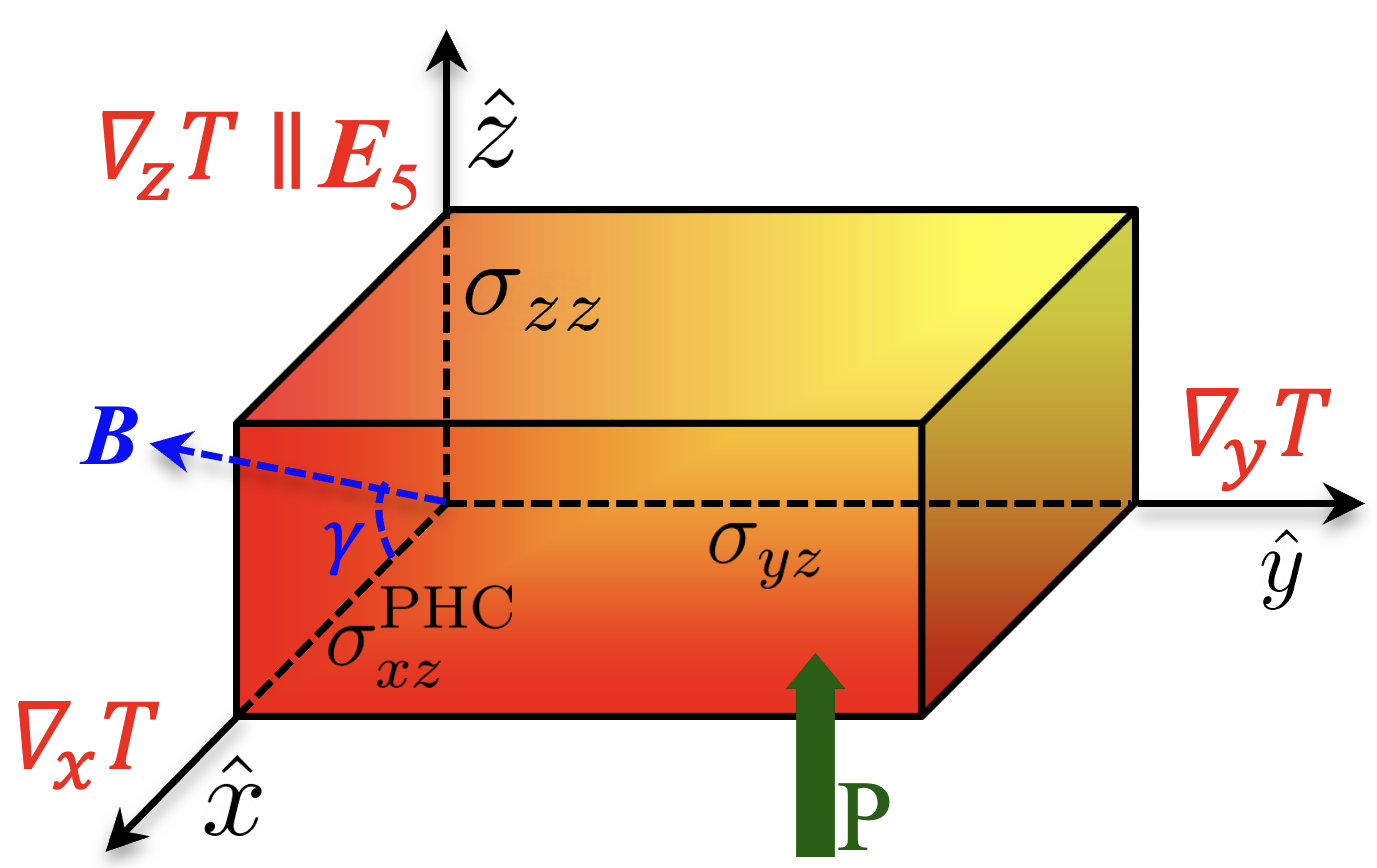}
    \caption {A schematic description of the strain-induced Ettingshausen effect. An inhomogeneous strain ($\mathbf{P}$) is applied in the z-direction that induces a chiral electric field $\mathbf{E_5}$ in the same direction. The magnetic field $\mathbf{B}$ is applied along the $xz$-plane, making an angle $\gamma$ with the $x$-axis. Depending on the angle $\gamma$, temperature gradients $\partial_x{T}$, $\partial_y{T}$, and $\partial_z{T}$ are induced in one or more  directions; $\sigma_{zz}$ represents the longitudinal magnetoconductance, $\sigma_{yz}$ arises due to the conventional Lorentz contribution, and $\sigma_{xz}$ is the chiral-anomaly induced planar-Hall conductivity.}
\label{Fig:schematic}
\end{figure}

Fig.~\ref{Fig:schematic} presents a schematic description of the experimental setup that we work with in this paper. An inhomogeneous strain is applied in the $z$-direction that induces a chiral electric field $\mathbf{E_5}$ in the same direction. The magnetic field $\mathbf{B}$ is applied along the $xz$-plane, making an angle $\gamma$ with the $x$-axis. In the absence of any charge current, which we impose as an experimental constraint, the generated temperature gradient can be found using the relation:
\begin{align}
  \hat{\sigma}\mathbf{E_5}=\hat{\alpha}\nabla{T}.
  \label{temp_grad_eqn}
\end{align}
Equation~\ref{temp_grad_eqn} is expanded as 
\begin{align}
  \begin{bmatrix}
\sigma_{xx} & \sigma_{xy} & \sigma_{xz} \\
\sigma_{yx} & \sigma_{yy} & \sigma_{yz} \\
\sigma_{zx} & \sigma_{zy} & \sigma_{zz}
\end{bmatrix}
\begin{bmatrix}
0 \\
0 \\
E_5
\end{bmatrix}
=
\begin{bmatrix}
\alpha_{xx} & \alpha_{xy} & \alpha_{xz} \\
\alpha_{yx} & \alpha_{yy} & \alpha_{yz} \\
\alpha_{zx} & \alpha_{zy} & \alpha_{zz}
\end{bmatrix}
\begin{bmatrix}
\partial_x T \\
\partial_y T \\
\partial_z T
\end{bmatrix}
\label{Eq_matrix} 
\end{align}
From the above equation, we can see that the components of the $\hat{\sigma}$ and $\hat{\alpha}$ matrices affect the generated temperature gradients ($\partial_xT$,$\partial_yT$,$\partial_zT$). Specifically, we note that the components $\sigma_{xz}$, $\sigma_{yz}$, and $\sigma_{zz}$ influence the generated temperature gradients, and these three components correspond to the CA-induced planar Hall conductivity, Lorentz-driven Hall conductivity, and the CA-driven longitudinal magnetoconductivity. 
In what follows, we discuss three interesting geometries to study the nature of the induced temperature gradients.

\subsubsection{Parallel electric and magnetic fields ($\gamma=\pi/2$)}
\begin{figure}
    \centering
    \includegraphics[width=.98\columnwidth]{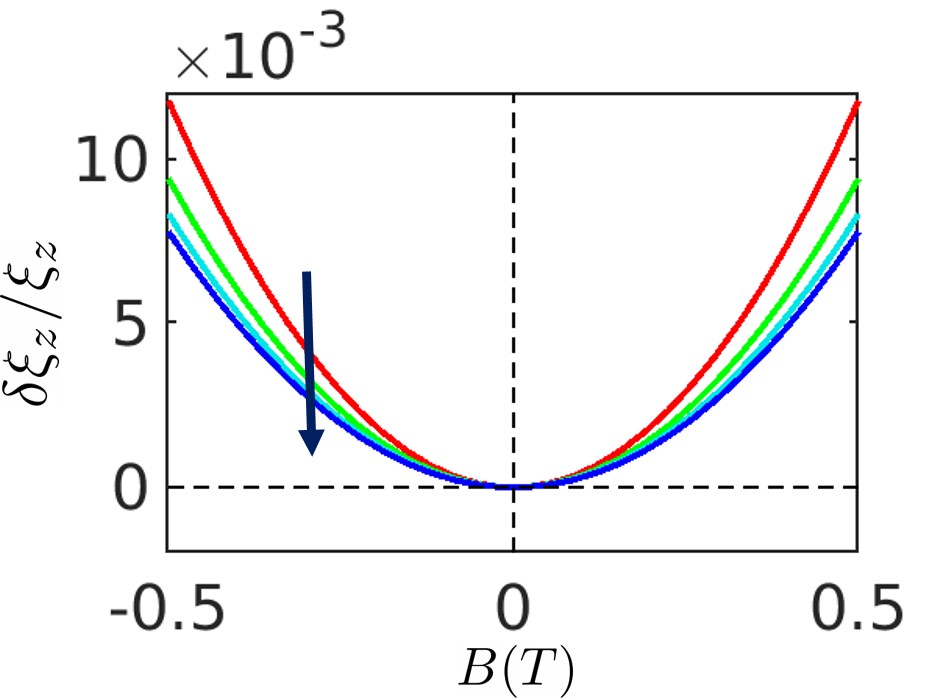}
    \caption {A  measure of the temperature gradient ($\xi_z = {\partial_z T}/{E_5}$) along $z$-direction when $\gamma=\pi/2$, i.e., the applied magnetic field and the strain-induced electric field are parallel. Different curves represent different values of the interband scattering $\alpha$. The arrow indicates the direction of increasing $\alpha$. Notably, there is no sign reversal for any arbitrary value of  $\alpha$. Note that  $\delta \xi_{z}/\xi_{z}=({\xi_{z}(B)-\xi_{z}(0)})/{\xi_{z}(0)}$.}
\label{TzE5_b_non_zero_gm_piby2_new}
\end{figure}

First, we study the geometry when the applied magnetic field is parallel to the strain-induced electric field and magnetic field, i.e., $\gamma=\pi/2$. In this case, we obtain that the temperature gradient is only along the $z$-direction, which is parallel to the applied magnetic field. As shown in Fig.~\ref{TzE5_b_non_zero_gm_piby2_new}, we can see that $\xi_z (\equiv\partial_zT/E_5)$ is quadratic in magnetic field. In the figure, we present the evaluated LMC for different values of the interband scattering $\alpha$ that range from $\alpha=0.35$ to $\alpha=0.9$. The value of $\alpha$ increases as we move from the red curve to the dark blue curve and the temperature gradient decreases. The quadratic nature of the curve can be explained as follows. Since $\mathbf{E_5}\parallel\mathbf{B}$, we only have contributions to $\xi_z$ from the longitudinal magnetoconductance $\sigma_{zz}$ and the Peltier coefficient $\alpha_{zz}$, and therefore, 
$\xi_z={\sigma_{zz}}/{\alpha_{zz}}$.
Previously, we demonstrated that SOC-NCMs exhibit a longitudinal electrical conductivity $\sigma_{zz}$ that shows a quadratic dependence on the applied magnetic field due to the chiral anomaly. The thermoelectric coefficient $\alpha_{zz}$ displays a similar quadratic response, which also originates from the chiral anomaly. However, we find that the relative magnitude of this quadratic dependence in $\alpha_{zz}$ is weaker compared to $\sigma_{zz}$, rendering its influence on the magnetic field dependence of $\xi_z$ negligible. Consequently, $\xi_z$ inherits a predominantly quadratic magnetic field dependence. Notably, neither $\sigma_{zz}$ nor $\alpha_{zz}$ exhibits any sign reversal across arbitrary values of the interband scattering rate, and as a result, $\xi_z$ doesn't exhibit any reversal of sign as well. This is in sharp contrast to Weyl semimetals, where a change of sign is observed beyond a critical intervalley scattering strength~\cite{knoll2020negative,sharma2023decoupling}. 
Additionally, the temperature gradient diminishes with increasing $\alpha$, reflecting the same trend observed in $\sigma_{zz}$. In this regime, the strain-induced Ettingshausen effect in SOC-NCMs is governed solely by contributions from the chiral anomaly.

\subsubsection{Electric field and magnetic fields are perpendicular, $\gamma=0$}
We next discuss the geometry where the applied magnetic field is perpendicular to the strain-induced electric field ($\gamma=0$). In this case, interestingly, we obtain temperature gradients along both the $y$- and $z$-directions due to the contributions from normal Hall conductance driven by the Lorentz force. The analytical expressions for the temperature gradients along $y$ and $z$ directions are given as follows:
\begin{align}
    \xi_y=\frac{\partial_y T}{E_5}=\frac{\alpha_{yz}\sigma_{zz}-\alpha_{zz}\sigma_{yz}} {\alpha_{yz}^{2}+\alpha_{yy}\alpha_{zz}}, \label{Eq:xi_y}\\
     \xi_z=\frac{\partial_z T}{E_5}=\frac{\alpha_{yy}\sigma_{zz}-\alpha_{zy}\sigma_{yz}}{\alpha_{yz}^{2}+\alpha_{yy}\alpha_{zz}} .\label{Eq:xi_z}
\end{align}
In Eq.~\ref{Eq:xi_y} and Eq.~\ref{Eq:xi_z},~$\alpha_{yz}$ and $\sigma_{yz}$ are the Lorentz contributions to the conductivities due to temperature gradient response and electric response, respectively, given by the following expressions:
\begin{align}
\sigma_{yz} 
= \notag \\ -&\frac{e^3B}{\hbar^2} \int{\frac{d^3\mathbf{k}}{(2\pi)^3}\tau_\mathbf{k}^2 \left(\frac{-\partial f}{\partial\epsilon_\mathbf{k}}\right)\left(\frac{v_y^2\partial^2\epsilon_\mathbf{k}}{\partial k_z^2}-\frac{v_yv_z\partial^2\epsilon_\mathbf{k}}{\partial k_y\partial k_z}\right)} \label{c} \\
  \alpha_{yz} 
  =\notag \\\frac{e^2B}{\hbar^2} &\int{\frac{d^3\mathbf{k}}{(2\pi)^3}\frac{(\epsilon-\mu)}{{T}}\tau_\mathbf{k}^2 \left(\frac{-\partial f}{\partial\epsilon_\mathbf{k}}\right)\left(\frac{v_y^2\partial^2\epsilon_\mathbf{k}}{\partial k_z^2}-\frac{v_yv_z\partial^2\epsilon_\mathbf{k}}{\partial k_y\partial k_z}\right)}. \label{d}
\end{align}
Here $(-{\partial f}/{\partial \epsilon_\mathbf{k}})$ is the derivative of the Fermi-Dirac distribution at a finite temperature ${T}$,~$v_y$ and $v_z$ are the velocities of the chiral fermions along the $y$- and $z$-directions, and $\tau_\mathbf{k}$ is the momentum dependent relaxation time (evaluated in Sec.~\ref{2d}). Note that since the electric and magnetic fields are orthogonal, the planar-Hall contributions due to the chiral anomaly vanish. 

\begin{figure}
    \centering
    \includegraphics[width=\columnwidth]{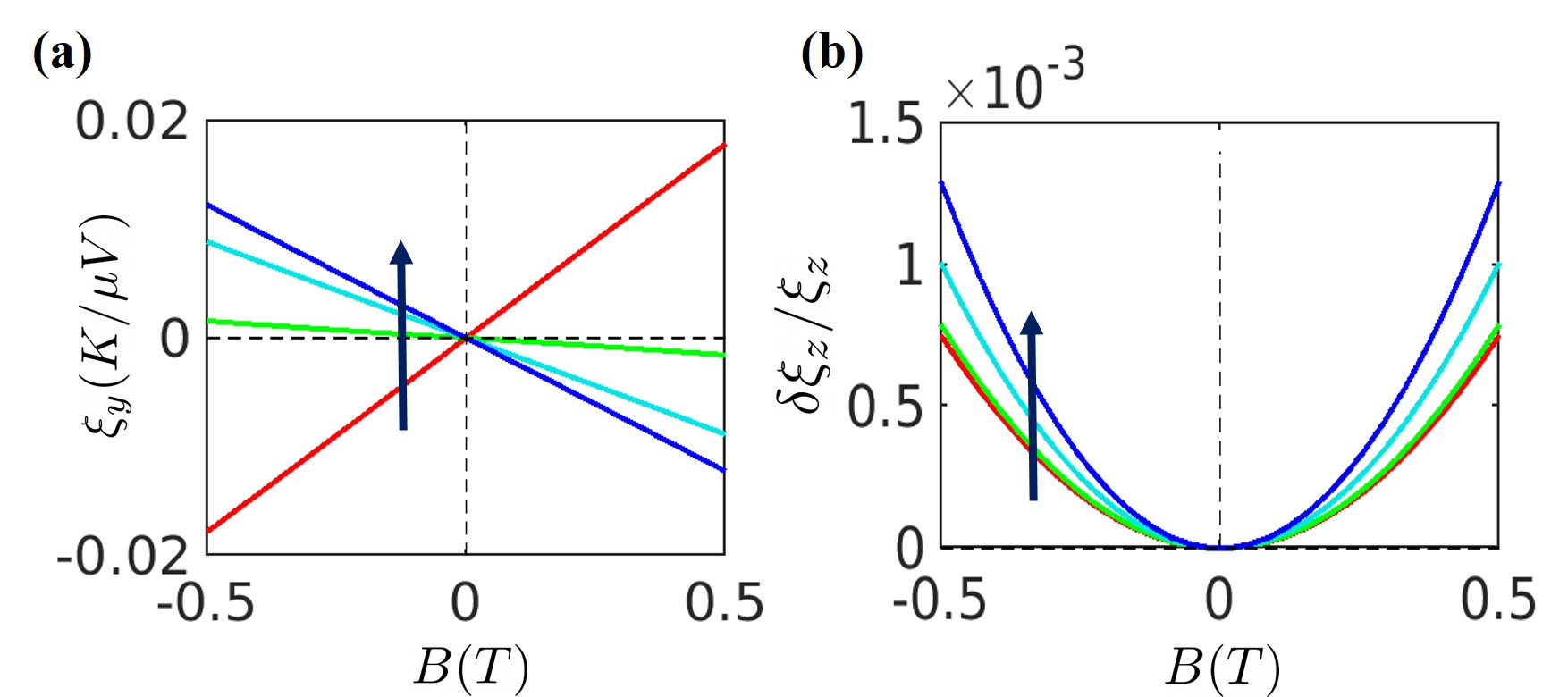}
    \caption {(a) Temperature gradient in the y-direction $\xi_y$ vs ${B}$ for different values of interband scattering $\alpha$ for $\gamma=0$. (b) A qualitative measure of the temperature gradient in the z-direction as a function of $B$. The arrow indicates the direction of increasing $\alpha$. It can be seen that $\xi_y$ flips sign after a certain value of $\alpha$. Note that $\xi_y=\frac{\partial_yT}{E_5}$ and $\xi_z=\frac{\partial_zT}{E_5}$.}
\label{TxE5_b_non_zero_gm_piby2}
\end{figure}

For the temperature gradient along the y-direction as shown in Fig.~\ref{TxE5_b_non_zero_gm_piby2}(a), we see that $\xi_y$ is linear in magnetic field. The arrow indicates the direction of increasing interband scattering. Here, we uncover an interesting result. The temperature gradient in the y-direction $\xi_y$ flips its sign when the interband scattering crosses a threshold. 
The temperature gradient along the $z$-direction ($\xi_z$), as shown in Fig.~\ref{TxE5_b_non_zero_gm_piby2}(b), is quadratic in the magnetic field, and unlike the former case, $\xi_z$ does not show sign reversal even for sufficiently strong interband scattering. 
The linear-in-${B}$ nature of $\xi_y$ is directly influenced by the linear-in-$B$ dependence of $\sigma_{yz}$ and $\alpha_{yz}$. The sign flip beyond a threshold value of interband scattering $\alpha$ is explained using Eq.~\ref{Eq:xi_y}. The two terms in the numerator have similar orders of magnitude and carry the same sign (this is closely related to Sondheimer's cancellation of the Nernst signal in a metal~\cite{sondheimer1948theory,wang2001onset,behnia2009nernst}); it is the competition between these two terms that results in the sign change as a function of interband scattering. In fact, it is well-known that, unlike the Lorentz-Hall and Peltier coefficients, the Nernst-Ettingshausen signal is not correlated with the sign of the carriers. Here, we uncover that the sign of the Ettingshausen signal is correlated with the strength of the interband scattering. 


The quadratic dependence of $\xi_z$ on ${B}$ is explained using Eq.~\ref{Eq:xi_z}. Due to the quadratic dependence of the Lorentz-force-driven conductivities on the scattering time $\tau_k$, their magnitudes are less compared to the chiral-anomaly induced longitudinal conductivities. As a result, the terms that involve solely Lorentz contributions are lesser in magnitude. Hence, the quadratic nature of chiral-anomaly-induced LMC dominates over the linear nature of Lorentz Hall, reflecting in the overall quadratic nature in $\xi_z$. It is also interesting to note that, in this case, $\xi_z$ increases with interband scattering $\alpha$, unlike for the case when $\gamma=\pi/2$. This feature is aided by the purely Lorentz term (second term in Eq.~\ref{Eq:xi_z}). For small $\alpha$, this term has a greater contribution, thereby reducing the overall magnitude of the numerator, reducing $\xi_z$. However, as $\alpha$ is increased, the second term decreases drastically in magnitude, thereby enhancing $\xi_z$. \\

\subsubsection{General case: $\mathbf{E_5}$ and $\mathbf{B}$ are at an arbitrary angle $\gamma$}

\begin{figure*}
    \centering
    \includegraphics[width=1.9\columnwidth]{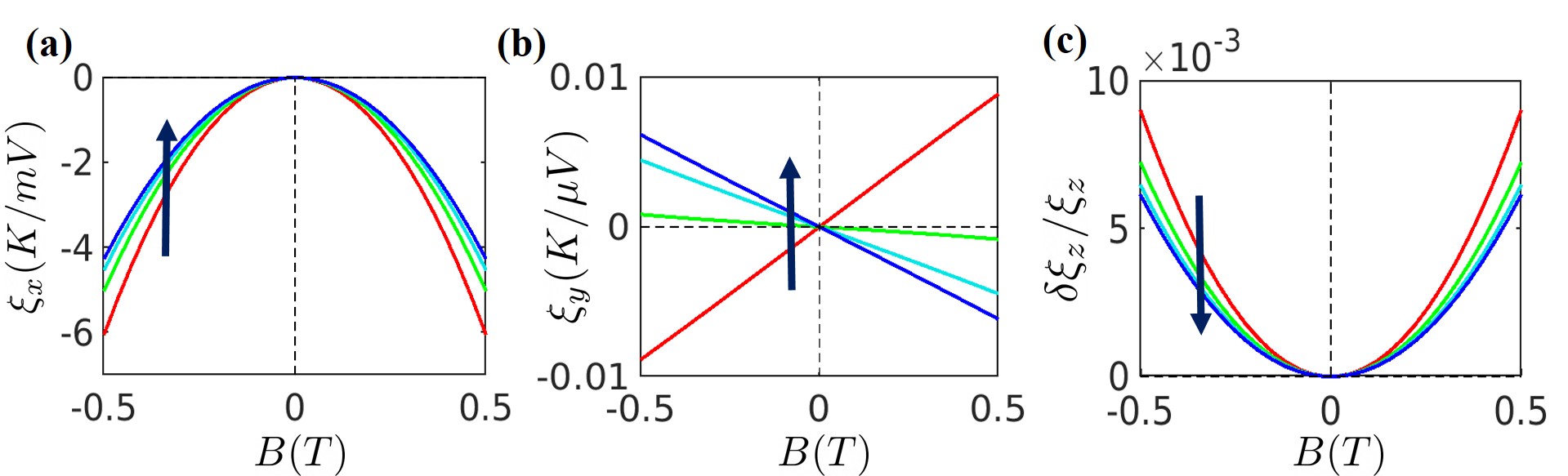}
    \caption {(a) Temperature gradient in the x-direction $\xi_x$ vs ${B}$ for different values of interband scattering $\alpha$ for $\gamma=\pi/3$. (b) Temperature gradient in the y-direction $\xi_y$ vs ${B}$ for different values of interband scattering $\alpha$ for $\gamma=\pi/3$.(c) A qualitative measure of the temperature gradient in the $z$-direction as a function of $B$; the arrow indicates the direction of increasing $\alpha$. It is seen that $\xi_y$ alone flips sign after a certain value of $\alpha$. Note that $\xi_x=\frac{\partial_xT}{E_5}$,$\xi_y=\frac{\partial_yT}{E_5}$ and $\xi_z=\frac{\partial_zT}{E_5}$.}
\label{TyE5_b_non_zero_gm_piby2}
\end{figure*}

In this case, we obtain temperature gradients along all three directions (see Fig.~\ref{TyE5_b_non_zero_gm_piby2}). All the components of the $\hat{\alpha}$ matrix in Eq.~\ref{Eq_matrix} become significant, because unlike the above two cases, all contributions come into the picture at the same time: the chiral-anomaly induced LMC ($\sigma_{zz}$), the chiral-anomaly induced planar-Hall conductivity ($\sigma_{xz}$) and the Lorentz-driven conductivity ($\sigma_{yz}$). 
As shown in Fig.~\ref{TyE5_b_non_zero_gm_piby2}(b), we see that the temperature gradient along the $y$-direction flips sign at a certain value of interband scattering. Similar to the previous case, this happens due to the competition between the CA-induced LMCs, PHCs, and the Lorentz contributions. $\xi_z$ exhibits the quadratic nature due to the combined effects of chiral-anomaly-induced LMCs and PHCs (Fig.~\ref{TyE5_b_non_zero_gm_piby2} (c)). The temperature gradient along the $x$-direction, $\xi_x$, is also quadratic in $\mathbf{B}$ (Fig.~\ref{TyE5_b_non_zero_gm_piby2} (a)). Strikingly, we obtain a \textit{negative} temperature gradient whose magnitude decreases with increasing interband scattering. We understand these features by first examining the expression for the temperature gradient along the $x$-direction, which is given by:
\begin{widetext}
\begin{align}
\xi_x=\frac{\partial_xT}{E_5}=\frac{\alpha_{yz}\alpha_{zy}\sigma_{xz}-\alpha_{yy}\alpha_{zz}\sigma_{xz}-\alpha_{xz}\alpha_{zy}\sigma_{yz}+\alpha_{xy}\alpha_{zz}\sigma_{yz}+\alpha_{xz}\alpha_{yy}\sigma_{zz}-\alpha_{xy}\alpha_{yz}\sigma_{zz}}{\alpha_{xz}\alpha_{yy}\alpha_{zx}-\alpha_{xy}\alpha_{yz}\alpha_{zx}-\alpha_{xz}\alpha_{yx}\alpha_{zy}+\alpha_{xx}\alpha_{yz}\alpha_{zy}+\alpha_{xy}\alpha_{yx}\alpha_{zz}-\alpha_{xx}\alpha_{yy}\alpha_{zz}}
\label{Ex}
\end{align}
\end{widetext}
From Eq.~\ref{Ex}, we note that there are three dominant terms, i.e., ones which are solely due to CA-induced LMCs and PHCs. These terms include the second and the second last terms of the numerator and the last term of the denominator. The rest of the terms have products involving Lorentz contributions, which greatly reduce their magnitudes and hence also their significant contribution to $\xi_x$. As a result, Eq.~\ref{Ex} can be approximated as $\xi_x\approx({\alpha_{xz}\sigma_{zz}-\alpha_{zz}\sigma_{xz}})/{\alpha_{xx}\alpha_{zz}}$. Here, too, we have two terms in the numerator, which are products of CA-induced LMCs and PHCs and have comparable orders of magnitude. Since PHC has a much dominant quadratic dependence than that of LMC, $\xi_x$ follows the dominant dependence of PHC. Since the second term ($\alpha_{zz}\sigma_{xz}$) comes with a sign that is the negative of the first term, it reflects itself in $\xi_x$. Here, we infer that the temperature gradient along the $x$-direction is largely dictated by PHC. 

\begin{figure*}
    \centering
    \includegraphics[width=1.9\columnwidth]{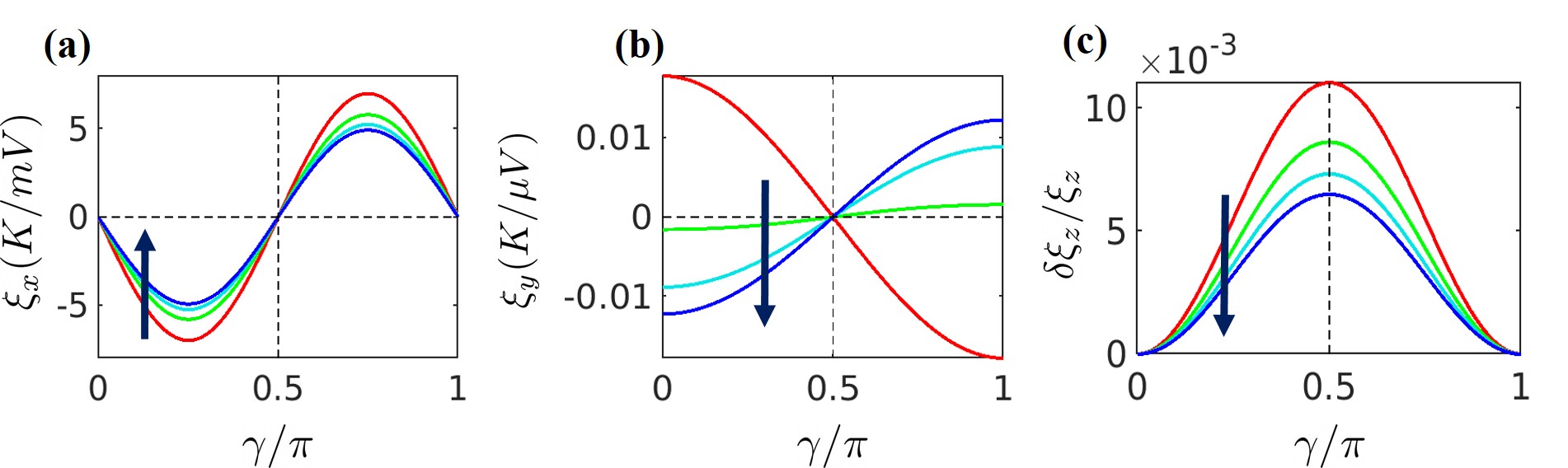}
    \caption{(a) Temperature gradient along $x$-direction $\xi_x$ as a function of $\gamma$. (b) Temperature gradient along $y$-direction $\xi_y$ vs $\gamma$ (c) Temperature gradient along $z$-direction $\xi_z$ vs $\gamma$. Here, we choose a fixed value of ${B}=0.5T$. Plot (a) shows a $\sin(2\gamma)$ behavior which is qualitatively similar to that of PHC. Plot (b) follows a $\sim\cos(\gamma$) curve, following the characteristics of the normal Hall effect. Plot (c) on the right shows a $\sin^2(\gamma$) behavior which is qualitatively similar to that of LMC.}
    \label{gamma_dep}
\end{figure*}
    

From the above discussion, it is evident that the direction and magnitude of the strain-induced temperature gradient can be altered by the angle $\gamma$ between ${E_5}$ and ${B}$ fields, and also the interband scattering. The full picture of how these parameters affect temperature gradients can be understood from Fig.~\ref{gamma_dep}, where temperature gradients are evaluated as a function of $\gamma$ for a fixed magnitude of ${B}$. In Fig.~\ref{gamma_dep}(a), we find that $\xi_x$ varies as $\sin(2\gamma)$, similar to the variation of chiral-anomaly-induced PHC~\cite{nandy2017chiral}-- showing that it is dictated by PHC. In Fig.~\ref{gamma_dep}(b), we see that $\xi_y$ becomes zero at $\gamma=\pi/2$ (as seen before) showing a $\cos(\gamma)$ trend. Following the arrow, we can see that there is a flip in sign for the temperature gradient along the $y$-direction as noted in our discussions above. In Fig.~\ref{gamma_dep}(c), we see that $\delta\xi_z/\xi_z$ shows a $\sin^2(\gamma$) trend that is qualitatively same as the $\gamma$ dependence of LMC (not plotted explicitly). This again highlights that $\xi_z$ is purely dictated by CA-induced LMC.
\subsection{Anomalous Ettingshausen effect in SOC-NCMs} 
Since SOC-NCMs break inversion symmetry but respect time-reversal symmetry, they do not essentially show Berry curvature-induced anomalous Hall effect. Time-reversal symmetry ensures that the integral of the Berry curvature weighted by the Boltzmann distribution function vanishes exactly.
However, in the presence of a magnetic field (which breaks TR symmetry), due to the coupling via orbital magnetic moment ($\epsilon_\mathbf{k}^\lambda\rightarrow \epsilon_\mathbf{k}^\lambda - \mathbf{m}^\lambda\cdot\mathbf{B}$), the two bands have asymmetric Fermi surfaces.  The Berry curvature thus contributes to a net anomalous Hall conductance~\cite{varma2024magnetotransport}. 
While the relative magnitudes of the Lorentz-induced conductivity and the Berry-curvature-induced conductivity can vary depending on factors such as overall disorder strength, interband scattering, and so on, here we focus solely on the effect of the latter. The Lorentz force-induced contributions are therefore neglected. The expressions for the anomalous Hall and Peltier contributions to conductivity are given below:
\begin{align}
\sigma^a_{yz}=\frac{e^2}{\hbar} \sum_{\lambda}\int \frac{d^3\mathbf{k}}{(2\pi)^3}~f_\mathbf{k}^\lambda\boldsymbol{\Omega}_\mathbf{k}^{\lambda,a},
\end{align}
\begin{align}
\alpha^{a}_{yz}=\frac{-k_Be}{\hbar}\sum_\lambda \int\frac{d^3\mathbf{k}}{(2\pi)^3}~s_\mathbf{k}^\lambda\boldsymbol{\Omega}_\mathbf{k}^{\lambda,a},
\end{align}
where $s_\mathbf{k}=-f_\mathbf{k}\log(f_\mathbf{k})-[(1-f_\mathbf{k})\log(1-f_\mathbf{k})]$ is the entropy density. Using these expressions along with Eq.~\ref{Eq_matrix} we evaluate the temperature gradients $\xi_i$. 
\begin{figure*}
    \centering
    \includegraphics[width=1.9\columnwidth]{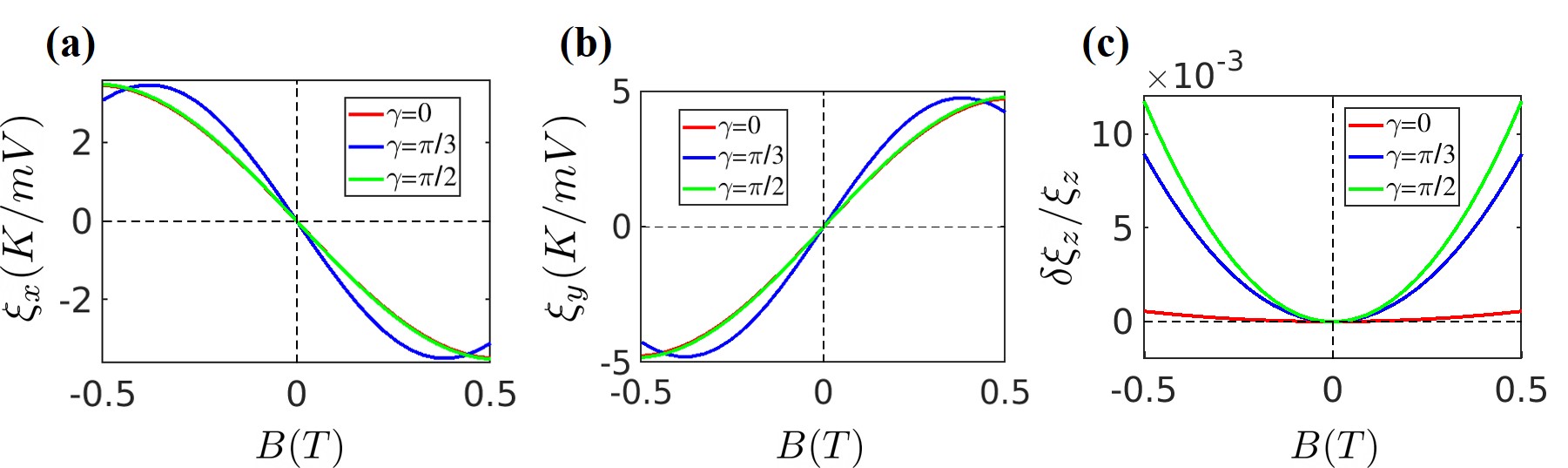}
    \caption{Temperature gradients along $x$, $y$ and $z$-directions as a function of magnetic field $B$, when AHE dominates over Lorentz contribution. We fix $\alpha=0.35$. Dependence on $\gamma$ is studied for three cases $\gamma=0,\pi/3$ and $\pi/2$.}
    
\label{gamma_anomalous}
\end{figure*}
As shown in Fig.~\ref{gamma_anomalous}(a) and Fig.~\ref{gamma_anomalous}(b), $\xi_x$ and $\xi_y$ have opposing non-monotonic dependence on ${B}$ while $\xi_z$ retains the quadratic dependence.  
The non-monotonous nature of the anomalous Hall conductivity is due to the combined effects of Berry curvature distribution and the OMM.
The opposite signs of temperature gradients in Fig.~\ref{gamma_anomalous}(a) and Fig.~\ref{gamma_anomalous}(b) stem from the opposite signs of the respective anomalous Hall conductivities that largely influence these temperature gradients. To be more precise, note that the anomalous Hall conductivities vary as $\mathbf{E_5}\times \mathbf{\Omega}$. Since, $\mathbf{E_5}$ acts along the z-direction,~$\sigma_{xz}^a$ and $\sigma_{yz}^a$ that depend on $\Omega_y$ and $\Omega_x$ respectively will have opposite signs. Now, 
$\xi_x$ is dictated by anomalous Hall $\sigma_{xz}^a$, which under sufficiently large interband scattering dominates over the planar Hall contribution $\sigma_{xz}$, whereas, $\xi_y$ is dictated by the anomalous Hall $\sigma_{yz}^a$, which by our assumption dominates over the Lorentz contribution $\sigma_{yz}$. 
In Fig.~\ref{gamma_anomalous}(c), which plots $\xi_z$, the effect of CA is retained in the quadratic dependence on $B$. This, as we have seen in earlier discussions, is due to the dominant nature of chiral anomaly induced LMC $(\sigma_{zz})$, which is quadratic in magnetic field.

It is also worthwhile to note the $\gamma$-dependence of the temperature gradients along the three directions. In Fig.~\ref{gamma_anomalous}(a) and Fig.~\ref{gamma_anomalous}(b), the $\gamma$-dependence arises from the effect of OMM through the term $\mathbf{m}\cdot\mathbf{B}$, and the fact that the angular variation of the magnetic field will result in an angular dependence of the components of the Berry curvature. However, in Fig.~\ref{gamma_anomalous}(c), the $\gamma$ dependence stems from the $\mathbf{E}\cdot\mathbf{B}$ term in CA. As $\gamma$ increases from $0$ to $\pi/2$~(red to green curve in figure), the contribution from the CA term increases, thereby increasing LMC. Since LMC drives the temperature gradient along the $z$-direction, we see that it increases correspondingly.
\begin{figure}
    \centering
    \includegraphics[width=\columnwidth]{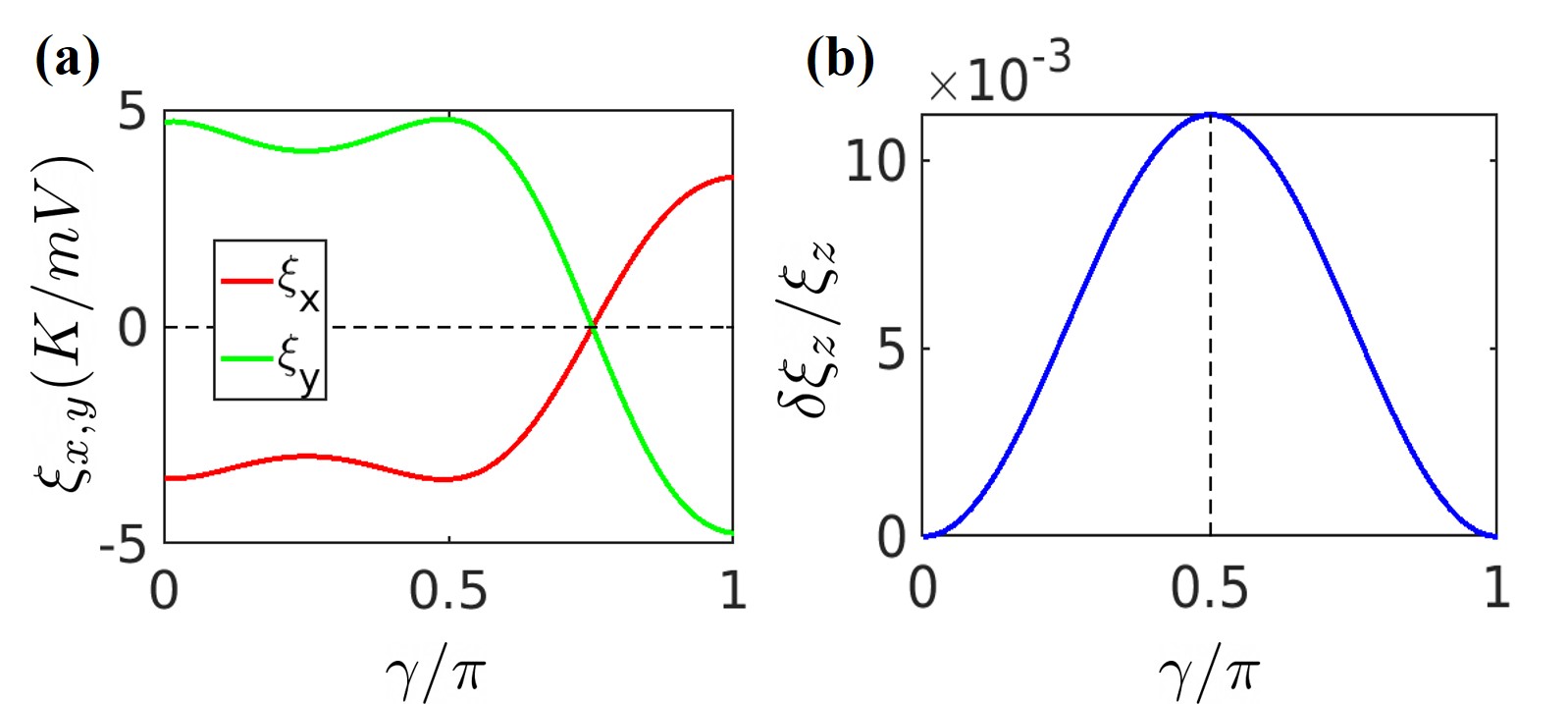}
    \caption{Temperature gradients along (a) $x$-direction $(\xi_x)$ and $y$-direction $(\xi_y)$, (b) $z$-direction$(\xi_z)$ are plotted as a function of angle $\gamma$. We fix the magnetic field $B=0.5T$.}
    
\label{gamma_anomalous_2}
\end{figure}
To better understand the anomalous Ettingshausen effect, we plot all three temperature gradients as a function of the angle $\gamma$ as shown in Fig.~\ref{gamma_anomalous_2}. Fig.~\ref{gamma_anomalous_2}(a) again shows that $\xi_x$ and $\xi_y$ are dictated by the anomalous Hall conductivities that have opposite signs. The non-monotonous nature arises due to the intricate dependency of the anomalous Hall on the OMM. Interestingly, we can also see that $\xi_x$ and $\xi_y$ reverse sign beyond a critical value of $\gamma$. At the critical value, the temperature gradients along the $x$ and $y$ directions vanish while retaining the one along the $z$-direction. As shown in Fig.~\ref{gamma_anomalous_2}(b), $\xi_z$ retains the $\sin^2(\gamma)$ dependence of the CA induced LMC. This shows that even if the anomalous Hall dominates over the Lorentz and planar Hall conductivities, the temperature gradient along the z-direction is always dictated by CA.

\section{Conclusion} \label{4}
In this paper, we examined the strain-induced Ettingshausen effect in spin-orbit-coupled noncentrosymmetric metals. Using a tight-binding lattice model under inhomogeneous strain, we find that strain induces anisotropy in the spin-orbit coupling. Furthermore, the inhomogeneous nature of the strain generates an axial electric field $\mathbf{E}_5$ in the direction of strain. This strain-induced field, in the presence of an external ${B}$-field, generates temperature gradients in three directions in the steady state.
 
As we explored the factors affecting these temperature gradients, we found that their presence in each direction is determined by the presence of the three dominating conductivities along the three directions--namely the longitudinal magnetoconductivity, Lorentz Hall and the planar Hall conductivity, which is in turn determined by the angle $\gamma$ between $\mathbf{E_5}$ and $\mathbf{B}$. We find that when $\gamma=\pi/2$, we only have temperature gradient along the $z$-direction (i.e., $\partial_zT$). Driven solely by chiral-anomaly-induced LMC, it has a quadratic dependence on ${B}$. When $\gamma=0$, we have a non-vanishing Lorentz Hall term, but a vanishing planar-Hall term, leading to temperature gradients along both the $y$- and $z$-directions (the $\partial_yT$ is determined by the presence of  Lorentz Hall). We also find that the temperature gradient along $y$-direction, $\partial_yT$, is linear in ${B}$, and also flips sign beyond a threshold value of interband scattering, while $\partial_zT$ does not. This is due to the different dependencies of the two conductivities on the interband scattering strength. For all other values of $\gamma$, we have finite temperature gradients along all three directions as we have finite conductivities along all three directions (longitudinal, planar-Hall, and Lorentz force induced). Here, $\partial_xT$ depends on the presence of CA-induced planar Hall conductivity. We also examined how the qualitative nature of the temperature gradients changes when Berry-curvature-driven anomalous Hall dominates over the Lorentz Hall and planar Hall contributions. 
We find that the magnitudes and signs of $\partial_xT$ and $\partial_yT$ can be tuned by varying the angle between $\mathbf{E_5}$ and $\mathbf{B}$, and that at a certain angle both the gradients vanish exactly. 
Our findings are highly relevant for upcoming experiments on these materials. By measuring strain-induced temperature gradients and examining their dependence on external magnetic fields, we can extract valuable insights about the tensorial structure and microscopic origins of the various conductivity components in the system. 
 
\section{Acknowledgements}
G.V.K acknowledges funding from IIT Mandi HTRA. A.A. acknowledges funding from IITM/ANRF-SERB/GS/495.  G.S. acknowledges funding from grant IITM/ANRF-SERB/GS/495. \\
\newpage
\appendix

\section{Low-energy expansion}
The lattice model in the main text can be expanded near the high symmetry points as shown in Eq.~\ref{Eq:low-E-expansion}. Note that to bring the spin-orbit coupling in the form $\chi\mathbf{k}\cdot\boldsymbol{\sigma}$, a suitable unitary transformation is performed in each case.  
\begin{widetext}
\begin{align}
    h_{0,0,0} &= 2t_1 \left( 3 - \frac{a^2{k}^2}{2} \right) 
    + 4t_2 (3 - a^2{k}^2) 
    + 8t_3 \left( 1 - \frac{a^2{k}^2}{2} \right) 
    + at_{\text{so}} \mathbf{k} \cdot \boldsymbol{\sigma}, \nonumber \\
    h_{\pi,\pi,\pi} &= -2t_1 \left( 3 - \frac{a^2{k}^2}{2} \right) 
    + 4t_2 (3 - a^2{k}^2) 
    - 8t_3 \left( 1 - \frac{a^2{k}^2}{2} \right) 
    - at_{\text{so}} \mathbf{k} \cdot \boldsymbol{\sigma}, \nonumber \\
    h_{\pi,\pi,0} &= -2t_1 \left( 1 - \frac{a^2{k}^2 - 2a^2k_z^2}{2} \right) 
    - 4t_2 (1 - a^2k_z^2) 
    + 8t_3 \left( 1 - \frac{a^2{k}^2}{2} \right) + at_{\text{so}} \mathbf{k} \cdot \boldsymbol{\sigma}, \nonumber\\
    h_{0,0,\pi} &= 2t_1 \left( 1 - \frac{a^2{k}^2 - 2a^2k_z^2}{2} \right) 
    - 4t_2 (1 - a^2k_z^2) 
    - 8t_3 \left( 1 - \frac{a^2{k}^2}{2} \right) - at_{\text{so}} \mathbf{k} \cdot \boldsymbol{\sigma}, \nonumber\\
    h_{\pi,0,\pi} &= -2t_1 \left( 1 - \frac{a^2{k}^2 - 2a^2k_y^2}{2} \right) 
    - 4t_2 (1 - a^2k_y^2) 
    + 8t_3 \left( 1 - \frac{a^2{k}^2}{2} \right)  + at_{\text{so}} \mathbf{k} \cdot \boldsymbol{\sigma}, \nonumber\\
    h_{0,\pi,0} &= 2t_1 \left( 1 - \frac{a^2{k}^2 - 2a^2k_y^2}{2} \right) 
    - 4t_2 (1 - a^2k_y^2) 
    - 8t_3 \left( 1 - \frac{a^2{k}^2}{2} \right)  - at_{\text{so}} \mathbf{k} \cdot \boldsymbol{\sigma}, \nonumber\\
    h_{0,\pi,\pi} &= -2t_1 \left( 1 - \frac{a^2{k}^2 - 2a^2k_x^2}{2} \right) 
    - 4t_2 (1 - a^2k_x^2) 
    + 8t_3 \left( 1 - \frac{a^2{k}^2}{2} \right)  + at_{\text{so}} \mathbf{k} \cdot \boldsymbol{\sigma}, \nonumber\\
    h_{\pi,0,0} &= 2t_1 \left( 1 - \frac{a^2{k}^2 - 2a^2k_x^2}{2} \right) 
    - 4t_2 (1 - a^2k_x^2) 
    - 8t_3 \left( 1 - \frac{a^2{k}^2}{2} \right)  - at_{\text{so}} \mathbf{k} \cdot \boldsymbol{\sigma}.
    \label{Eq:low-E-expansion}
\end{align}
\end{widetext}

\pagebreak
 
\bibliography{biblio.bib}

\begin{thebibliography}{57}%
\makeatletter
\providecommand \@ifxundefined [1]{%
 \@ifx{#1\undefined}
}%
\providecommand \@ifnum [1]{%
 \ifnum #1\expandafter \@firstoftwo
 \else \expandafter \@secondoftwo
 \fi
}%
\providecommand \@ifx [1]{%
 \ifx #1\expandafter \@firstoftwo
 \else \expandafter \@secondoftwo
 \fi
}%
\providecommand \natexlab [1]{#1}%
\providecommand \enquote  [1]{``#1''}%
\providecommand \bibnamefont  [1]{#1}%
\providecommand \bibfnamefont [1]{#1}%
\providecommand \citenamefont [1]{#1}%
\providecommand \href@noop [0]{\@secondoftwo}%
\providecommand \href [0]{\begingroup \@sanitize@url \@href}%
\providecommand \@href[1]{\@@startlink{#1}\@@href}%
\providecommand \@@href[1]{\endgroup#1\@@endlink}%
\providecommand \@sanitize@url [0]{\catcode `\\12\catcode `\$12\catcode `\&12\catcode `\#12\catcode `\^12\catcode `\_12\catcode `\%12\relax}%
\providecommand \@@startlink[1]{}%
\providecommand \@@endlink[0]{}%
\providecommand \url  [0]{\begingroup\@sanitize@url \@url }%
\providecommand \@url [1]{\endgroup\@href {#1}{\urlprefix }}%
\providecommand \urlprefix  [0]{URL }%
\providecommand \Eprint [0]{\href }%
\providecommand \doibase [0]{https://doi.org/}%
\providecommand \selectlanguage [0]{\@gobble}%
\providecommand \bibinfo  [0]{\@secondoftwo}%
\providecommand \bibfield  [0]{\@secondoftwo}%
\providecommand \translation [1]{[#1]}%
\providecommand \BibitemOpen [0]{}%
\providecommand \bibitemStop [0]{}%
\providecommand \bibitemNoStop [0]{.\EOS\space}%
\providecommand \EOS [0]{\spacefactor3000\relax}%
\providecommand \BibitemShut  [1]{\csname bibitem#1\endcsname}%
\let\auto@bib@innerbib\@empty
\bibitem [{\citenamefont {Manes}(2007)}]{manes2007symmetry}%
  \BibitemOpen
  \bibfield  {author} {\bibinfo {author} {\bibfnamefont {J.~L.}\ \bibnamefont {Manes}},\ }\bibfield  {title} {\bibinfo {title} {Symmetry-based approach to electron-phonon interactions in graphene},\ }\href@noop {} {\bibfield  {journal} {\bibinfo  {journal} {Physical Review B—Condensed Matter and Materials Physics}\ }\textbf {\bibinfo {volume} {76}},\ \bibinfo {pages} {045430} (\bibinfo {year} {2007})}\BibitemShut {NoStop}%
\bibitem [{\citenamefont {Manes}\ \emph {et~al.}(2013)\citenamefont {Manes}, \citenamefont {de~Juan}, \citenamefont {Sturla},\ and\ \citenamefont {Vozmediano}}]{manes2013generalized}%
  \BibitemOpen
  \bibfield  {author} {\bibinfo {author} {\bibfnamefont {J.~L.}\ \bibnamefont {Manes}}, \bibinfo {author} {\bibfnamefont {F.}~\bibnamefont {de~Juan}}, \bibinfo {author} {\bibfnamefont {M.}~\bibnamefont {Sturla}},\ and\ \bibinfo {author} {\bibfnamefont {M.~A.}\ \bibnamefont {Vozmediano}},\ }\bibfield  {title} {\bibinfo {title} {Generalized effective hamiltonian for graphene under nonuniform strain},\ }\href@noop {} {\bibfield  {journal} {\bibinfo  {journal} {Physical Review B—Condensed Matter and Materials Physics}\ }\textbf {\bibinfo {volume} {88}},\ \bibinfo {pages} {155405} (\bibinfo {year} {2013})}\BibitemShut {NoStop}%
\bibitem [{\citenamefont {Iorio}\ and\ \citenamefont {Pais}(2015)}]{iorio2015revisiting}%
  \BibitemOpen
  \bibfield  {author} {\bibinfo {author} {\bibfnamefont {A.}~\bibnamefont {Iorio}}\ and\ \bibinfo {author} {\bibfnamefont {P.}~\bibnamefont {Pais}},\ }\bibfield  {title} {\bibinfo {title} {Revisiting the gauge fields of strained graphene},\ }\href@noop {} {\bibfield  {journal} {\bibinfo  {journal} {Physical Review D}\ }\textbf {\bibinfo {volume} {92}},\ \bibinfo {pages} {125005} (\bibinfo {year} {2015})}\BibitemShut {NoStop}%
\bibitem [{\citenamefont {Sohier}\ \emph {et~al.}(2014)\citenamefont {Sohier}, \citenamefont {Calandra}, \citenamefont {Park}, \citenamefont {Bonini}, \citenamefont {Marzari},\ and\ \citenamefont {Mauri}}]{sohier2014phonon}%
  \BibitemOpen
  \bibfield  {author} {\bibinfo {author} {\bibfnamefont {T.}~\bibnamefont {Sohier}}, \bibinfo {author} {\bibfnamefont {M.}~\bibnamefont {Calandra}}, \bibinfo {author} {\bibfnamefont {C.-H.}\ \bibnamefont {Park}}, \bibinfo {author} {\bibfnamefont {N.}~\bibnamefont {Bonini}}, \bibinfo {author} {\bibfnamefont {N.}~\bibnamefont {Marzari}},\ and\ \bibinfo {author} {\bibfnamefont {F.}~\bibnamefont {Mauri}},\ }\bibfield  {title} {\bibinfo {title} {Phonon-limited resistivity of graphene by first-principles calculations: Electron-phonon interactions, strain-induced gauge field, and boltzmann equation},\ }\href@noop {} {\bibfield  {journal} {\bibinfo  {journal} {Physical Review B}\ }\textbf {\bibinfo {volume} {90}},\ \bibinfo {pages} {125414} (\bibinfo {year} {2014})}\BibitemShut {NoStop}%
\bibitem [{\citenamefont {Vozmediano}\ \emph {et~al.}(2010)\citenamefont {Vozmediano}, \citenamefont {Katsnelson},\ and\ \citenamefont {Guinea}}]{vozmediano2010gauge}%
  \BibitemOpen
  \bibfield  {author} {\bibinfo {author} {\bibfnamefont {M.~A.}\ \bibnamefont {Vozmediano}}, \bibinfo {author} {\bibfnamefont {M.}~\bibnamefont {Katsnelson}},\ and\ \bibinfo {author} {\bibfnamefont {F.}~\bibnamefont {Guinea}},\ }\bibfield  {title} {\bibinfo {title} {Gauge fields in graphene},\ }\href@noop {} {\bibfield  {journal} {\bibinfo  {journal} {Physics Reports}\ }\textbf {\bibinfo {volume} {496}},\ \bibinfo {pages} {109} (\bibinfo {year} {2010})}\BibitemShut {NoStop}%
\bibitem [{\citenamefont {Castro~Neto}\ \emph {et~al.}(2009)\citenamefont {Castro~Neto}, \citenamefont {Guinea}, \citenamefont {Peres}, \citenamefont {Novoselov},\ and\ \citenamefont {Geim}}]{castro2009electronic}%
  \BibitemOpen
  \bibfield  {author} {\bibinfo {author} {\bibfnamefont {A.~H.}\ \bibnamefont {Castro~Neto}}, \bibinfo {author} {\bibfnamefont {F.}~\bibnamefont {Guinea}}, \bibinfo {author} {\bibfnamefont {N.~M.}\ \bibnamefont {Peres}}, \bibinfo {author} {\bibfnamefont {K.~S.}\ \bibnamefont {Novoselov}},\ and\ \bibinfo {author} {\bibfnamefont {A.~K.}\ \bibnamefont {Geim}},\ }\bibfield  {title} {\bibinfo {title} {The electronic properties of graphene},\ }\href@noop {} {\bibfield  {journal} {\bibinfo  {journal} {Reviews of modern physics}\ }\textbf {\bibinfo {volume} {81}},\ \bibinfo {pages} {109} (\bibinfo {year} {2009})}\BibitemShut {NoStop}%
\bibitem [{\citenamefont {Low}\ and\ \citenamefont {Guinea}(2010)}]{low2010strain}%
  \BibitemOpen
  \bibfield  {author} {\bibinfo {author} {\bibfnamefont {T.}~\bibnamefont {Low}}\ and\ \bibinfo {author} {\bibfnamefont {F.}~\bibnamefont {Guinea}},\ }\bibfield  {title} {\bibinfo {title} {Strain-induced pseudomagnetic field for novel graphene electronics},\ }\href@noop {} {\bibfield  {journal} {\bibinfo  {journal} {Nano letters}\ }\textbf {\bibinfo {volume} {10}},\ \bibinfo {pages} {3551} (\bibinfo {year} {2010})}\BibitemShut {NoStop}%
\bibitem [{\citenamefont {Guinea}\ \emph {et~al.}(2010)\citenamefont {Guinea}, \citenamefont {Katsnelson},\ and\ \citenamefont {Geim}}]{guineaenergy}%
  \BibitemOpen
  \bibfield  {author} {\bibinfo {author} {\bibfnamefont {F.}~\bibnamefont {Guinea}}, \bibinfo {author} {\bibfnamefont {M.}~\bibnamefont {Katsnelson}},\ and\ \bibinfo {author} {\bibfnamefont {A.}~\bibnamefont {Geim}},\ }\bibfield  {title} {\bibinfo {title} {Energy gaps, topological insulator state and zero-field quantum hall effect in graphene by strain engineering},\ }\href@noop {} {\bibfield  {journal} {\bibinfo  {journal} {Nature Physics}\ }\textbf {\bibinfo {volume} {6}},\ \bibinfo {pages} {30} (\bibinfo {year} {2010})}\BibitemShut {NoStop}%
\bibitem [{\citenamefont {Settnes}\ \emph {et~al.}(2017)\citenamefont {Settnes}, \citenamefont {Garcia},\ and\ \citenamefont {Roche}}]{settnes2017valley}%
  \BibitemOpen
  \bibfield  {author} {\bibinfo {author} {\bibfnamefont {M.}~\bibnamefont {Settnes}}, \bibinfo {author} {\bibfnamefont {J.~H.}\ \bibnamefont {Garcia}},\ and\ \bibinfo {author} {\bibfnamefont {S.}~\bibnamefont {Roche}},\ }\bibfield  {title} {\bibinfo {title} {Valley-polarized quantum transport generated by gauge fields in graphene},\ }\href@noop {} {\bibfield  {journal} {\bibinfo  {journal} {2D Materials}\ }\textbf {\bibinfo {volume} {4}},\ \bibinfo {pages} {031006} (\bibinfo {year} {2017})}\BibitemShut {NoStop}%
\bibitem [{\citenamefont {Cortijo}\ \emph {et~al.}(2015)\citenamefont {Cortijo}, \citenamefont {Ferreir{\'o}s}, \citenamefont {Landsteiner},\ and\ \citenamefont {Vozmediano}}]{cortijo2015elastic}%
  \BibitemOpen
  \bibfield  {author} {\bibinfo {author} {\bibfnamefont {A.}~\bibnamefont {Cortijo}}, \bibinfo {author} {\bibfnamefont {Y.}~\bibnamefont {Ferreir{\'o}s}}, \bibinfo {author} {\bibfnamefont {K.}~\bibnamefont {Landsteiner}},\ and\ \bibinfo {author} {\bibfnamefont {M.~A.}\ \bibnamefont {Vozmediano}},\ }\bibfield  {title} {\bibinfo {title} {Elastic gauge fields in weyl semimetals},\ }\href@noop {} {\bibfield  {journal} {\bibinfo  {journal} {Physical review letters}\ }\textbf {\bibinfo {volume} {115}},\ \bibinfo {pages} {177202} (\bibinfo {year} {2015})}\BibitemShut {NoStop}%
\bibitem [{\citenamefont {Cortijo}\ \emph {et~al.}(2016)\citenamefont {Cortijo}, \citenamefont {Kharzeev}, \citenamefont {Landsteiner},\ and\ \citenamefont {Vozmediano}}]{cortijo2016strain}%
  \BibitemOpen
  \bibfield  {author} {\bibinfo {author} {\bibfnamefont {A.}~\bibnamefont {Cortijo}}, \bibinfo {author} {\bibfnamefont {D.}~\bibnamefont {Kharzeev}}, \bibinfo {author} {\bibfnamefont {K.}~\bibnamefont {Landsteiner}},\ and\ \bibinfo {author} {\bibfnamefont {M.~A.}\ \bibnamefont {Vozmediano}},\ }\bibfield  {title} {\bibinfo {title} {Strain-induced chiral magnetic effect in weyl semimetals},\ }\href@noop {} {\bibfield  {journal} {\bibinfo  {journal} {Physical Review B}\ }\textbf {\bibinfo {volume} {94}},\ \bibinfo {pages} {241405} (\bibinfo {year} {2016})}\BibitemShut {NoStop}%
\bibitem [{\citenamefont {Shapourian}\ \emph {et~al.}(2015)\citenamefont {Shapourian}, \citenamefont {Hughes},\ and\ \citenamefont {Ryu}}]{shapourian2015viscoelastic}%
  \BibitemOpen
  \bibfield  {author} {\bibinfo {author} {\bibfnamefont {H.}~\bibnamefont {Shapourian}}, \bibinfo {author} {\bibfnamefont {T.~L.}\ \bibnamefont {Hughes}},\ and\ \bibinfo {author} {\bibfnamefont {S.}~\bibnamefont {Ryu}},\ }\bibfield  {title} {\bibinfo {title} {Viscoelastic response of topological tight-binding models in two and three dimensions},\ }\href@noop {} {\bibfield  {journal} {\bibinfo  {journal} {Physical Review B}\ }\textbf {\bibinfo {volume} {92}},\ \bibinfo {pages} {165131} (\bibinfo {year} {2015})}\BibitemShut {NoStop}%
\bibitem [{\citenamefont {Burkov}(2014)}]{burkov2014anomalous}%
  \BibitemOpen
  \bibfield  {author} {\bibinfo {author} {\bibfnamefont {A.}~\bibnamefont {Burkov}},\ }\bibfield  {title} {\bibinfo {title} {Anomalous hall effect in weyl metals},\ }\href@noop {} {\bibfield  {journal} {\bibinfo  {journal} {Physical review letters}\ }\textbf {\bibinfo {volume} {113}},\ \bibinfo {pages} {187202} (\bibinfo {year} {2014})}\BibitemShut {NoStop}%
\bibitem [{\citenamefont {Sharma}\ \emph {et~al.}(2016)\citenamefont {Sharma}, \citenamefont {Goswami},\ and\ \citenamefont {Tewari}}]{sharma2016nernst}%
  \BibitemOpen
  \bibfield  {author} {\bibinfo {author} {\bibfnamefont {G.}~\bibnamefont {Sharma}}, \bibinfo {author} {\bibfnamefont {P.}~\bibnamefont {Goswami}},\ and\ \bibinfo {author} {\bibfnamefont {S.}~\bibnamefont {Tewari}},\ }\bibfield  {title} {\bibinfo {title} {Nernst and magnetothermal conductivity in a lattice model of weyl fermions},\ }\href@noop {} {\bibfield  {journal} {\bibinfo  {journal} {Physical Review B}\ }\textbf {\bibinfo {volume} {93}},\ \bibinfo {pages} {035116} (\bibinfo {year} {2016})}\BibitemShut {NoStop}%
\bibitem [{\citenamefont {Sharma}\ \emph {et~al.}(2017)\citenamefont {Sharma}, \citenamefont {Moore}, \citenamefont {Saha},\ and\ \citenamefont {Tewari}}]{sharma2017nernst}%
  \BibitemOpen
  \bibfield  {author} {\bibinfo {author} {\bibfnamefont {G.}~\bibnamefont {Sharma}}, \bibinfo {author} {\bibfnamefont {C.}~\bibnamefont {Moore}}, \bibinfo {author} {\bibfnamefont {S.}~\bibnamefont {Saha}},\ and\ \bibinfo {author} {\bibfnamefont {S.}~\bibnamefont {Tewari}},\ }\bibfield  {title} {\bibinfo {title} {Nernst effect in dirac and inversion-asymmetric weyl semimetals},\ }\href@noop {} {\bibfield  {journal} {\bibinfo  {journal} {Physical Review B}\ }\textbf {\bibinfo {volume} {96}},\ \bibinfo {pages} {195119} (\bibinfo {year} {2017})}\BibitemShut {NoStop}%
\bibitem [{\citenamefont {Wan}\ \emph {et~al.}(2011)\citenamefont {Wan}, \citenamefont {Turner}, \citenamefont {Vishwanath},\ and\ \citenamefont {Savrasov}}]{wan2011topological}%
  \BibitemOpen
  \bibfield  {author} {\bibinfo {author} {\bibfnamefont {X.}~\bibnamefont {Wan}}, \bibinfo {author} {\bibfnamefont {A.~M.}\ \bibnamefont {Turner}}, \bibinfo {author} {\bibfnamefont {A.}~\bibnamefont {Vishwanath}},\ and\ \bibinfo {author} {\bibfnamefont {S.~Y.}\ \bibnamefont {Savrasov}},\ }\bibfield  {title} {\bibinfo {title} {Topological semimetal and fermi-arc surface states in the electronic structure of pyrochlore iridates},\ }\href@noop {} {\bibfield  {journal} {\bibinfo  {journal} {Physical Review B—Condensed Matter and Materials Physics}\ }\textbf {\bibinfo {volume} {83}},\ \bibinfo {pages} {205101} (\bibinfo {year} {2011})}\BibitemShut {NoStop}%
\bibitem [{\citenamefont {Nandy}\ \emph {et~al.}(2017)\citenamefont {Nandy}, \citenamefont {Sharma}, \citenamefont {Taraphder},\ and\ \citenamefont {Tewari}}]{nandy2017chiral}%
  \BibitemOpen
  \bibfield  {author} {\bibinfo {author} {\bibfnamefont {S.}~\bibnamefont {Nandy}}, \bibinfo {author} {\bibfnamefont {G.}~\bibnamefont {Sharma}}, \bibinfo {author} {\bibfnamefont {A.}~\bibnamefont {Taraphder}},\ and\ \bibinfo {author} {\bibfnamefont {S.}~\bibnamefont {Tewari}},\ }\bibfield  {title} {\bibinfo {title} {Chiral anomaly as the origin of the planar hall effect in weyl semimetals},\ }\href@noop {} {\bibfield  {journal} {\bibinfo  {journal} {Physical review letters}\ }\textbf {\bibinfo {volume} {119}},\ \bibinfo {pages} {176804} (\bibinfo {year} {2017})}\BibitemShut {NoStop}%
\bibitem [{\citenamefont {Adler}(1969)}]{adler1969axial}%
  \BibitemOpen
  \bibfield  {author} {\bibinfo {author} {\bibfnamefont {S.~L.}\ \bibnamefont {Adler}},\ }\bibfield  {title} {\bibinfo {title} {Axial-vector vertex in spinor electrodynamics},\ }\href@noop {} {\bibfield  {journal} {\bibinfo  {journal} {Physical Review}\ }\textbf {\bibinfo {volume} {177}},\ \bibinfo {pages} {2426} (\bibinfo {year} {1969})}\BibitemShut {NoStop}%
\bibitem [{\citenamefont {Aji}(2012)}]{aji2012adler}%
  \BibitemOpen
  \bibfield  {author} {\bibinfo {author} {\bibfnamefont {V.}~\bibnamefont {Aji}},\ }\bibfield  {title} {\bibinfo {title} {Adler-bell-jackiw anomaly in weyl semimetals: Application to pyrochlore iridates},\ }\href@noop {} {\bibfield  {journal} {\bibinfo  {journal} {Physical Review B—Condensed Matter and Materials Physics}\ }\textbf {\bibinfo {volume} {85}},\ \bibinfo {pages} {241101} (\bibinfo {year} {2012})}\BibitemShut {NoStop}%
\bibitem [{\citenamefont {Nielsen}\ and\ \citenamefont {Ninomiya}(1981)}]{nielsen1981no}%
  \BibitemOpen
  \bibfield  {author} {\bibinfo {author} {\bibfnamefont {H.~B.}\ \bibnamefont {Nielsen}}\ and\ \bibinfo {author} {\bibfnamefont {M.}~\bibnamefont {Ninomiya}},\ }\href@noop {} {\emph {\bibinfo {title} {No-go theorum for regularizing chiral fermions}}},\ \bibinfo {type} {Tech. Rep.}\ (\bibinfo  {institution} {Science Research Council},\ \bibinfo {year} {1981})\BibitemShut {NoStop}%
\bibitem [{\citenamefont {Nielsen}\ and\ \citenamefont {Ninomiya}(1983)}]{nielsen1983adler}%
  \BibitemOpen
  \bibfield  {author} {\bibinfo {author} {\bibfnamefont {H.~B.}\ \bibnamefont {Nielsen}}\ and\ \bibinfo {author} {\bibfnamefont {M.}~\bibnamefont {Ninomiya}},\ }\bibfield  {title} {\bibinfo {title} {The adler-bell-jackiw anomaly and weyl fermions in a crystal},\ }\href@noop {} {\bibfield  {journal} {\bibinfo  {journal} {Physics Letters B}\ }\textbf {\bibinfo {volume} {130}},\ \bibinfo {pages} {389} (\bibinfo {year} {1983})}\BibitemShut {NoStop}%
\bibitem [{\citenamefont {Bell}\ and\ \citenamefont {Jackiw}(1969)}]{bell1969pcac}%
  \BibitemOpen
  \bibfield  {author} {\bibinfo {author} {\bibfnamefont {J.~S.}\ \bibnamefont {Bell}}\ and\ \bibinfo {author} {\bibfnamefont {R.}~\bibnamefont {Jackiw}},\ }\bibfield  {title} {\bibinfo {title} {A pcac puzzle: in the model},\ }\href@noop {} {\bibfield  {journal} {\bibinfo  {journal} {Il Nuovo Cimento A}\ }\textbf {\bibinfo {volume} {60}},\ \bibinfo {pages} {47} (\bibinfo {year} {1969})}\BibitemShut {NoStop}%
\bibitem [{\citenamefont {Goswami}\ and\ \citenamefont {Tewari}(2013{\natexlab{a}})}]{goswami2013axionic}%
  \BibitemOpen
  \bibfield  {author} {\bibinfo {author} {\bibfnamefont {P.}~\bibnamefont {Goswami}}\ and\ \bibinfo {author} {\bibfnamefont {S.}~\bibnamefont {Tewari}},\ }\bibfield  {title} {\bibinfo {title} {Axionic field theory of (3+ 1)-dimensional weyl semimetals},\ }\href@noop {} {\bibfield  {journal} {\bibinfo  {journal} {Physical Review B—Condensed Matter and Materials Physics}\ }\textbf {\bibinfo {volume} {88}},\ \bibinfo {pages} {245107} (\bibinfo {year} {2013}{\natexlab{a}})}\BibitemShut {NoStop}%
\bibitem [{\citenamefont {Goswami}\ \emph {et~al.}(2015)\citenamefont {Goswami}, \citenamefont {Sharma},\ and\ \citenamefont {Tewari}}]{goswami2015optical}%
  \BibitemOpen
  \bibfield  {author} {\bibinfo {author} {\bibfnamefont {P.}~\bibnamefont {Goswami}}, \bibinfo {author} {\bibfnamefont {G.}~\bibnamefont {Sharma}},\ and\ \bibinfo {author} {\bibfnamefont {S.}~\bibnamefont {Tewari}},\ }\bibfield  {title} {\bibinfo {title} {Optical activity as a test for dynamic chiral magnetic effect of weyl semimetals},\ }\href@noop {} {\bibfield  {journal} {\bibinfo  {journal} {Physical Review B}\ }\textbf {\bibinfo {volume} {92}},\ \bibinfo {pages} {161110} (\bibinfo {year} {2015})}\BibitemShut {NoStop}%
\bibitem [{\citenamefont {Goswami}\ and\ \citenamefont {Tewari}(2013{\natexlab{b}})}]{goswami2013chiral}%
  \BibitemOpen
  \bibfield  {author} {\bibinfo {author} {\bibfnamefont {P.}~\bibnamefont {Goswami}}\ and\ \bibinfo {author} {\bibfnamefont {S.}~\bibnamefont {Tewari}},\ }\bibfield  {title} {\bibinfo {title} {Chiral magnetic effect of weyl fermions and its applications to cubic noncentrosymmetric metals},\ }\href@noop {} {\bibfield  {journal} {\bibinfo  {journal} {arXiv preprint arXiv:1311.1506}\ } (\bibinfo {year} {2013}{\natexlab{b}})}\BibitemShut {NoStop}%
\bibitem [{\citenamefont {Fukushima}\ \emph {et~al.}(2008)\citenamefont {Fukushima}, \citenamefont {Kharzeev},\ and\ \citenamefont {Warringa}}]{fukushima2008chiral}%
  \BibitemOpen
  \bibfield  {author} {\bibinfo {author} {\bibfnamefont {K.}~\bibnamefont {Fukushima}}, \bibinfo {author} {\bibfnamefont {D.~E.}\ \bibnamefont {Kharzeev}},\ and\ \bibinfo {author} {\bibfnamefont {H.~J.}\ \bibnamefont {Warringa}},\ }\bibfield  {title} {\bibinfo {title} {Chiral magnetic effect},\ }\href@noop {} {\bibfield  {journal} {\bibinfo  {journal} {Physical Review D—Particles, Fields, Gravitation, and Cosmology}\ }\textbf {\bibinfo {volume} {78}},\ \bibinfo {pages} {074033} (\bibinfo {year} {2008})}\BibitemShut {NoStop}%
\bibitem [{\citenamefont {Ahmad}\ and\ \citenamefont {Sharma}(2021)}]{ahmad2021longitudinal}%
  \BibitemOpen
  \bibfield  {author} {\bibinfo {author} {\bibfnamefont {A.}~\bibnamefont {Ahmad}}\ and\ \bibinfo {author} {\bibfnamefont {G.}~\bibnamefont {Sharma}},\ }\bibfield  {title} {\bibinfo {title} {Longitudinal magnetoconductance and the planar hall effect in a lattice model of tilted weyl fermions},\ }\href@noop {} {\bibfield  {journal} {\bibinfo  {journal} {Physical Review B}\ }\textbf {\bibinfo {volume} {103}},\ \bibinfo {pages} {115146} (\bibinfo {year} {2021})}\BibitemShut {NoStop}%
\bibitem [{\citenamefont {Ahmad}\ \emph {et~al.}(2023)\citenamefont {Ahmad}, \citenamefont {Raman}, \citenamefont {Tewari},\ and\ \citenamefont {Sharma}}]{ahmad2023longitudinal}%
  \BibitemOpen
  \bibfield  {author} {\bibinfo {author} {\bibfnamefont {A.}~\bibnamefont {Ahmad}}, \bibinfo {author} {\bibfnamefont {K.~V.}\ \bibnamefont {Raman}}, \bibinfo {author} {\bibfnamefont {S.}~\bibnamefont {Tewari}},\ and\ \bibinfo {author} {\bibfnamefont {G.}~\bibnamefont {Sharma}},\ }\bibfield  {title} {\bibinfo {title} {Longitudinal magnetoconductance and the planar hall conductance in inhomogeneous weyl semimetals},\ }\href@noop {} {\bibfield  {journal} {\bibinfo  {journal} {Physical Review B}\ }\textbf {\bibinfo {volume} {107}},\ \bibinfo {pages} {144206} (\bibinfo {year} {2023})}\BibitemShut {NoStop}%
\bibitem [{\citenamefont {Ahmad}\ \emph {et~al.}(2024)\citenamefont {Ahmad}, \citenamefont {Varma},\ and\ \citenamefont {Sharma}}]{ahmad2024geometry}%
  \BibitemOpen
  \bibfield  {author} {\bibinfo {author} {\bibfnamefont {A.}~\bibnamefont {Ahmad}}, \bibinfo {author} {\bibfnamefont {G.}~\bibnamefont {Varma}},\ and\ \bibinfo {author} {\bibfnamefont {G.}~\bibnamefont {Sharma}},\ }\bibfield  {title} {\bibinfo {title} {Geometry, anomaly, topology, and transport in weyl fermions},\ }\href@noop {} {\bibfield  {journal} {\bibinfo  {journal} {Journal of Physics: Condensed Matter}\ }\textbf {\bibinfo {volume} {37}},\ \bibinfo {pages} {043001} (\bibinfo {year} {2024})}\BibitemShut {NoStop}%
\bibitem [{\citenamefont {Ahmad}\ and\ \citenamefont {Sharma}(2025)}]{ahmad2025longitudinal}%
  \BibitemOpen
  \bibfield  {author} {\bibinfo {author} {\bibfnamefont {A.}~\bibnamefont {Ahmad}}\ and\ \bibinfo {author} {\bibfnamefont {G.}~\bibnamefont {Sharma}},\ }\bibfield  {title} {\bibinfo {title} {Longitudinal magnetoconductance of higher-pseudospin fermions},\ }\href@noop {} {\bibfield  {journal} {\bibinfo  {journal} {Physical Review B}\ }\textbf {\bibinfo {volume} {112}},\ \bibinfo {pages} {045135} (\bibinfo {year} {2025})}\BibitemShut {NoStop}%
\bibitem [{\citenamefont {Varma~K}\ \emph {et~al.}(2025)\citenamefont {Varma~K}, \citenamefont {Raza},\ and\ \citenamefont {Ahmad}}]{Varma2025chiral}%
  \BibitemOpen
  \bibfield  {author} {\bibinfo {author} {\bibfnamefont {G.}~\bibnamefont {Varma~K}}, \bibinfo {author} {\bibfnamefont {M.}~\bibnamefont {Raza}},\ and\ \bibinfo {author} {\bibfnamefont {A.}~\bibnamefont {Ahmad}},\ }\bibfield  {title} {\bibinfo {title} {Chiral anomaly-induced nonlinear hall effect in spin-orbit coupled noncentrosymmetric metals},\ }\href@noop {} {\bibfield  {journal} {\bibinfo  {journal} {arXiv preprint arXiv:2508.00821}\ } (\bibinfo {year} {2025})}\BibitemShut {NoStop}%
\bibitem [{\citenamefont {Grushin}\ \emph {et~al.}(2016)\citenamefont {Grushin}, \citenamefont {Venderbos}, \citenamefont {Vishwanath},\ and\ \citenamefont {Ilan}}]{grushin2016inhomogeneous}%
  \BibitemOpen
  \bibfield  {author} {\bibinfo {author} {\bibfnamefont {A.~G.}\ \bibnamefont {Grushin}}, \bibinfo {author} {\bibfnamefont {J.~W.}\ \bibnamefont {Venderbos}}, \bibinfo {author} {\bibfnamefont {A.}~\bibnamefont {Vishwanath}},\ and\ \bibinfo {author} {\bibfnamefont {R.}~\bibnamefont {Ilan}},\ }\bibfield  {title} {\bibinfo {title} {Inhomogeneous weyl and dirac semimetals: Transport in axial magnetic fields and fermi arc surface states from pseudo-landau levels},\ }\href@noop {} {\bibfield  {journal} {\bibinfo  {journal} {Physical Review X}\ }\textbf {\bibinfo {volume} {6}},\ \bibinfo {pages} {041046} (\bibinfo {year} {2016})}\BibitemShut {NoStop}%
\bibitem [{\citenamefont {Ghosh}\ \emph {et~al.}(2020)\citenamefont {Ghosh}, \citenamefont {Sinha}, \citenamefont {Nandy},\ and\ \citenamefont {Taraphder}}]{ghosh2020chirality}%
  \BibitemOpen
  \bibfield  {author} {\bibinfo {author} {\bibfnamefont {S.}~\bibnamefont {Ghosh}}, \bibinfo {author} {\bibfnamefont {D.}~\bibnamefont {Sinha}}, \bibinfo {author} {\bibfnamefont {S.}~\bibnamefont {Nandy}},\ and\ \bibinfo {author} {\bibfnamefont {A.}~\bibnamefont {Taraphder}},\ }\bibfield  {title} {\bibinfo {title} {Chirality-dependent planar hall effect in inhomogeneous weyl semimetals},\ }\href@noop {} {\bibfield  {journal} {\bibinfo  {journal} {Physical Review B}\ }\textbf {\bibinfo {volume} {102}},\ \bibinfo {pages} {121105} (\bibinfo {year} {2020})}\BibitemShut {NoStop}%
\bibitem [{\citenamefont {Cheon}\ \emph {et~al.}(2022)\citenamefont {Cheon}, \citenamefont {Cho}, \citenamefont {Kim},\ and\ \citenamefont {Lee}}]{cheon2022chiral}%
  \BibitemOpen
  \bibfield  {author} {\bibinfo {author} {\bibfnamefont {S.}~\bibnamefont {Cheon}}, \bibinfo {author} {\bibfnamefont {G.~Y.}\ \bibnamefont {Cho}}, \bibinfo {author} {\bibfnamefont {K.-S.}\ \bibnamefont {Kim}},\ and\ \bibinfo {author} {\bibfnamefont {H.-W.}\ \bibnamefont {Lee}},\ }\bibfield  {title} {\bibinfo {title} {Chiral anomaly in noncentrosymmetric systems induced by spin-orbit coupling},\ }\href@noop {} {\bibfield  {journal} {\bibinfo  {journal} {Physical Review B}\ }\textbf {\bibinfo {volume} {105}},\ \bibinfo {pages} {L180303} (\bibinfo {year} {2022})}\BibitemShut {NoStop}%
\bibitem [{\citenamefont {Varma~K}\ \emph {et~al.}(2024)\citenamefont {Varma~K}, \citenamefont {Ahmad}, \citenamefont {Tewari},\ and\ \citenamefont {Sharma}}]{varma2024magnetotransport}%
  \BibitemOpen
  \bibfield  {author} {\bibinfo {author} {\bibfnamefont {G.}~\bibnamefont {Varma~K}}, \bibinfo {author} {\bibfnamefont {A.}~\bibnamefont {Ahmad}}, \bibinfo {author} {\bibfnamefont {S.}~\bibnamefont {Tewari}},\ and\ \bibinfo {author} {\bibfnamefont {G.}~\bibnamefont {Sharma}},\ }\bibfield  {title} {\bibinfo {title} {Magnetotransport in spin-orbit coupled noncentrosymmetric and weyl metals},\ }\href@noop {} {\bibfield  {journal} {\bibinfo  {journal} {Physical Review B}\ }\textbf {\bibinfo {volume} {109}},\ \bibinfo {pages} {165114} (\bibinfo {year} {2024})}\BibitemShut {NoStop}%
\bibitem [{\citenamefont {Ahmad}\ \emph {et~al.}(2025)\citenamefont {Ahmad}, \citenamefont {K},\ and\ \citenamefont {Sharma}}]{ahmad2025chiral}%
  \BibitemOpen
  \bibfield  {author} {\bibinfo {author} {\bibfnamefont {A.}~\bibnamefont {Ahmad}}, \bibinfo {author} {\bibfnamefont {G.~V.}\ \bibnamefont {K}},\ and\ \bibinfo {author} {\bibfnamefont {G.}~\bibnamefont {Sharma}},\ }\bibfield  {title} {\bibinfo {title} {Chiral anomaly induced nonlinear hall effect in three-dimensional chiral fermions},\ }\href@noop {} {\bibfield  {journal} {\bibinfo  {journal} {Physical Review B}\ }\textbf {\bibinfo {volume} {111}},\ \bibinfo {pages} {035138} (\bibinfo {year} {2025})}\BibitemShut {NoStop}%
\bibitem [{\citenamefont {Das}\ \emph {et~al.}(2023)\citenamefont {Das}, \citenamefont {Das},\ and\ \citenamefont {Agarwal}}]{das2023chiral}%
  \BibitemOpen
  \bibfield  {author} {\bibinfo {author} {\bibfnamefont {S.}~\bibnamefont {Das}}, \bibinfo {author} {\bibfnamefont {K.}~\bibnamefont {Das}},\ and\ \bibinfo {author} {\bibfnamefont {A.}~\bibnamefont {Agarwal}},\ }\bibfield  {title} {\bibinfo {title} {Chiral anomalies in three-dimensional spin-orbit coupled metals: Electrical, thermal, and gravitational anomalies},\ }\href@noop {} {\bibfield  {journal} {\bibinfo  {journal} {Physical Review B}\ }\textbf {\bibinfo {volume} {108}},\ \bibinfo {pages} {045405} (\bibinfo {year} {2023})}\BibitemShut {NoStop}%
\bibitem [{\citenamefont {Gopalakrishnan}\ \emph {et~al.}(2025)\citenamefont {Gopalakrishnan}, \citenamefont {Roy}, \citenamefont {Sharma},\ and\ \citenamefont {Tewari}}]{gopalakrishnan2025chiral}%
  \BibitemOpen
  \bibfield  {author} {\bibinfo {author} {\bibfnamefont {R.~G.}\ \bibnamefont {Gopalakrishnan}}, \bibinfo {author} {\bibfnamefont {B.~B.}\ \bibnamefont {Roy}}, \bibinfo {author} {\bibfnamefont {G.}~\bibnamefont {Sharma}},\ and\ \bibinfo {author} {\bibfnamefont {S.}~\bibnamefont {Tewari}},\ }\bibfield  {title} {\bibinfo {title} {Chiral anomaly induced transverse planar transport phenomena in three dimensional spin-orbit coupled metals},\ }\href@noop {} {\bibfield  {journal} {\bibinfo  {journal} {arXiv preprint arXiv:2505.21498}\ } (\bibinfo {year} {2025})}\BibitemShut {NoStop}%
\bibitem [{\citenamefont {Ri}\ \emph {et~al.}(1994)\citenamefont {Ri}, \citenamefont {Gross}, \citenamefont {Gollnik}, \citenamefont {Beck}, \citenamefont {Huebener}, \citenamefont {Wagner},\ and\ \citenamefont {Adrian}}]{ri1994nernst}%
  \BibitemOpen
  \bibfield  {author} {\bibinfo {author} {\bibfnamefont {H.-C.}\ \bibnamefont {Ri}}, \bibinfo {author} {\bibfnamefont {R.}~\bibnamefont {Gross}}, \bibinfo {author} {\bibfnamefont {F.}~\bibnamefont {Gollnik}}, \bibinfo {author} {\bibfnamefont {A.}~\bibnamefont {Beck}}, \bibinfo {author} {\bibfnamefont {R.}~\bibnamefont {Huebener}}, \bibinfo {author} {\bibfnamefont {P.}~\bibnamefont {Wagner}},\ and\ \bibinfo {author} {\bibfnamefont {H.}~\bibnamefont {Adrian}},\ }\bibfield  {title} {\bibinfo {title} {Nernst, seebeck, and hall effects in the mixed state of yba 2 cu 3 o 7- $\delta$ and bi 2 sr 2 cacu 2 o 8+ x thin films: A comparative study},\ }\href@noop {} {\bibfield  {journal} {\bibinfo  {journal} {Physical Review B}\ }\textbf {\bibinfo {volume} {50}},\ \bibinfo {pages} {3312} (\bibinfo {year} {1994})}\BibitemShut {NoStop}%
\bibitem [{\citenamefont {Behnia}\ and\ \citenamefont {Aubin}(2016)}]{behnia2016nernst}%
  \BibitemOpen
  \bibfield  {author} {\bibinfo {author} {\bibfnamefont {K.}~\bibnamefont {Behnia}}\ and\ \bibinfo {author} {\bibfnamefont {H.}~\bibnamefont {Aubin}},\ }\bibfield  {title} {\bibinfo {title} {Nernst effect in metals and superconductors: a review of concepts and experiments},\ }\href@noop {} {\bibfield  {journal} {\bibinfo  {journal} {Reports on Progress in Physics}\ }\textbf {\bibinfo {volume} {79}},\ \bibinfo {pages} {046502} (\bibinfo {year} {2016})}\BibitemShut {NoStop}%
\bibitem [{\citenamefont {Behnia}\ \emph {et~al.}(2007)\citenamefont {Behnia}, \citenamefont {M{\'e}asson},\ and\ \citenamefont {Kopelevich}}]{behnia2007nernst}%
  \BibitemOpen
  \bibfield  {author} {\bibinfo {author} {\bibfnamefont {K.}~\bibnamefont {Behnia}}, \bibinfo {author} {\bibfnamefont {M.-A.}\ \bibnamefont {M{\'e}asson}},\ and\ \bibinfo {author} {\bibfnamefont {Y.}~\bibnamefont {Kopelevich}},\ }\bibfield  {title} {\bibinfo {title} {Nernst effect in semimetals: The effective mass and the figure of merit},\ }\href@noop {} {\bibfield  {journal} {\bibinfo  {journal} {Physical review letters}\ }\textbf {\bibinfo {volume} {98}},\ \bibinfo {pages} {076603} (\bibinfo {year} {2007})}\BibitemShut {NoStop}%
\bibitem [{\citenamefont {Cyr-Choiniere}\ \emph {et~al.}(2009)\citenamefont {Cyr-Choiniere}, \citenamefont {Daou}, \citenamefont {Lalibert{\'e}}, \citenamefont {LeBoeuf}, \citenamefont {Doiron-Leyraud}, \citenamefont {Chang}, \citenamefont {Yan}, \citenamefont {Cheng}, \citenamefont {Zhou}, \citenamefont {Goodenough} \emph {et~al.}}]{cyr2009enhancement}%
  \BibitemOpen
  \bibfield  {author} {\bibinfo {author} {\bibfnamefont {O.}~\bibnamefont {Cyr-Choiniere}}, \bibinfo {author} {\bibfnamefont {R.}~\bibnamefont {Daou}}, \bibinfo {author} {\bibfnamefont {F.}~\bibnamefont {Lalibert{\'e}}}, \bibinfo {author} {\bibfnamefont {D.}~\bibnamefont {LeBoeuf}}, \bibinfo {author} {\bibfnamefont {N.}~\bibnamefont {Doiron-Leyraud}}, \bibinfo {author} {\bibfnamefont {J.}~\bibnamefont {Chang}}, \bibinfo {author} {\bibfnamefont {J.-Q.}\ \bibnamefont {Yan}}, \bibinfo {author} {\bibfnamefont {J.-G.}\ \bibnamefont {Cheng}}, \bibinfo {author} {\bibfnamefont {J.-S.}\ \bibnamefont {Zhou}}, \bibinfo {author} {\bibfnamefont {J.}~\bibnamefont {Goodenough}}, \emph {et~al.},\ }\bibfield  {title} {\bibinfo {title} {Enhancement of the nernst effect by stripe order in a high-t c superconductor},\ }\href@noop {} {\bibfield  {journal} {\bibinfo  {journal} {Nature}\ }\textbf {\bibinfo {volume} {458}},\ \bibinfo {pages} {743} (\bibinfo {year} {2009})}\BibitemShut {NoStop}%
\bibitem [{\citenamefont {Tinh}\ and\ \citenamefont {Rosenstein}(2009)}]{tinh2009theory}%
  \BibitemOpen
  \bibfield  {author} {\bibinfo {author} {\bibfnamefont {B.~D.}\ \bibnamefont {Tinh}}\ and\ \bibinfo {author} {\bibfnamefont {B.}~\bibnamefont {Rosenstein}},\ }\bibfield  {title} {\bibinfo {title} {Theory of nernst effect in high-t c superconductors},\ }\href@noop {} {\bibfield  {journal} {\bibinfo  {journal} {Physical Review B—Condensed Matter and Materials Physics}\ }\textbf {\bibinfo {volume} {79}},\ \bibinfo {pages} {024518} (\bibinfo {year} {2009})}\BibitemShut {NoStop}%
\bibitem [{\citenamefont {Wang}\ \emph {et~al.}(2006)\citenamefont {Wang}, \citenamefont {Li},\ and\ \citenamefont {Ong}}]{wang2006nernst}%
  \BibitemOpen
  \bibfield  {author} {\bibinfo {author} {\bibfnamefont {Y.}~\bibnamefont {Wang}}, \bibinfo {author} {\bibfnamefont {L.}~\bibnamefont {Li}},\ and\ \bibinfo {author} {\bibfnamefont {N.}~\bibnamefont {Ong}},\ }\bibfield  {title} {\bibinfo {title} {Nernst effect in high-t c superconductors},\ }\href@noop {} {\bibfield  {journal} {\bibinfo  {journal} {Physical Review B—Condensed Matter and Materials Physics}\ }\textbf {\bibinfo {volume} {73}},\ \bibinfo {pages} {024510} (\bibinfo {year} {2006})}\BibitemShut {NoStop}%
\bibitem [{\citenamefont {Liang}\ \emph {et~al.}(2017)\citenamefont {Liang}, \citenamefont {Lin}, \citenamefont {Gibson}, \citenamefont {Gao}, \citenamefont {Hirschberger}, \citenamefont {Liu}, \citenamefont {Cava},\ and\ \citenamefont {Ong}}]{liang2017anomalous}%
  \BibitemOpen
  \bibfield  {author} {\bibinfo {author} {\bibfnamefont {T.}~\bibnamefont {Liang}}, \bibinfo {author} {\bibfnamefont {J.}~\bibnamefont {Lin}}, \bibinfo {author} {\bibfnamefont {Q.}~\bibnamefont {Gibson}}, \bibinfo {author} {\bibfnamefont {T.}~\bibnamefont {Gao}}, \bibinfo {author} {\bibfnamefont {M.}~\bibnamefont {Hirschberger}}, \bibinfo {author} {\bibfnamefont {M.}~\bibnamefont {Liu}}, \bibinfo {author} {\bibfnamefont {R.~J.}\ \bibnamefont {Cava}},\ and\ \bibinfo {author} {\bibfnamefont {N.~P.}\ \bibnamefont {Ong}},\ }\bibfield  {title} {\bibinfo {title} {Anomalous nernst effect in the dirac semimetal cd 3 as 2},\ }\href@noop {} {\bibfield  {journal} {\bibinfo  {journal} {Physical review letters}\ }\textbf {\bibinfo {volume} {118}},\ \bibinfo {pages} {136601} (\bibinfo {year} {2017})}\BibitemShut {NoStop}%
\bibitem [{\citenamefont {Zhou}\ \emph {et~al.}(2022)\citenamefont {Zhou}, \citenamefont {Liu}, \citenamefont {Wu}, \citenamefont {Jiang}, \citenamefont {Shi}, \citenamefont {Li}, \citenamefont {Sui}, \citenamefont {Hu},\ and\ \citenamefont {Luo}}]{zhou2022anomalous}%
  \BibitemOpen
  \bibfield  {author} {\bibinfo {author} {\bibfnamefont {X.}~\bibnamefont {Zhou}}, \bibinfo {author} {\bibfnamefont {H.}~\bibnamefont {Liu}}, \bibinfo {author} {\bibfnamefont {W.}~\bibnamefont {Wu}}, \bibinfo {author} {\bibfnamefont {K.}~\bibnamefont {Jiang}}, \bibinfo {author} {\bibfnamefont {Y.}~\bibnamefont {Shi}}, \bibinfo {author} {\bibfnamefont {Z.}~\bibnamefont {Li}}, \bibinfo {author} {\bibfnamefont {Y.}~\bibnamefont {Sui}}, \bibinfo {author} {\bibfnamefont {J.}~\bibnamefont {Hu}},\ and\ \bibinfo {author} {\bibfnamefont {J.}~\bibnamefont {Luo}},\ }\bibfield  {title} {\bibinfo {title} {Anomalous thermal hall effect and anomalous nernst effect of csv 3 sb 5},\ }\href@noop {} {\bibfield  {journal} {\bibinfo  {journal} {Physical Review B}\ }\textbf {\bibinfo {volume} {105}},\ \bibinfo {pages} {205104} (\bibinfo {year} {2022})}\BibitemShut {NoStop}%
\bibitem [{\citenamefont {Roychowdhury}\ \emph {et~al.}(2023)\citenamefont {Roychowdhury}, \citenamefont {Yao}, \citenamefont {Samanta}, \citenamefont {Bae}, \citenamefont {Chen}, \citenamefont {Ju}, \citenamefont {Raghavan}, \citenamefont {Kumar}, \citenamefont {Constantinou}, \citenamefont {Guin} \emph {et~al.}}]{roychowdhury2023anomalous}%
  \BibitemOpen
  \bibfield  {author} {\bibinfo {author} {\bibfnamefont {S.}~\bibnamefont {Roychowdhury}}, \bibinfo {author} {\bibfnamefont {M.}~\bibnamefont {Yao}}, \bibinfo {author} {\bibfnamefont {K.}~\bibnamefont {Samanta}}, \bibinfo {author} {\bibfnamefont {S.}~\bibnamefont {Bae}}, \bibinfo {author} {\bibfnamefont {D.}~\bibnamefont {Chen}}, \bibinfo {author} {\bibfnamefont {S.}~\bibnamefont {Ju}}, \bibinfo {author} {\bibfnamefont {A.}~\bibnamefont {Raghavan}}, \bibinfo {author} {\bibfnamefont {N.}~\bibnamefont {Kumar}}, \bibinfo {author} {\bibfnamefont {P.}~\bibnamefont {Constantinou}}, \bibinfo {author} {\bibfnamefont {S.~N.}\ \bibnamefont {Guin}}, \emph {et~al.},\ }\bibfield  {title} {\bibinfo {title} {Anomalous hall conductivity and nernst effect of the ideal weyl semimetallic ferromagnet eucd2as2},\ }\href@noop {} {\bibfield  {journal} {\bibinfo  {journal} {Advanced Science}\ }\textbf {\bibinfo {volume} {10}},\ \bibinfo {pages} {2207121} (\bibinfo {year} {2023})}\BibitemShut {NoStop}%
\bibitem [{\citenamefont {Mukherjee}\ and\ \citenamefont {Takimoto}(2012)}]{mukherjee2012order}%
  \BibitemOpen
  \bibfield  {author} {\bibinfo {author} {\bibfnamefont {S.~P.}\ \bibnamefont {Mukherjee}}\ and\ \bibinfo {author} {\bibfnamefont {T.}~\bibnamefont {Takimoto}},\ }\bibfield  {title} {\bibinfo {title} {Order parameter with line nodes and s$\pm$-wave symmetry for the noncentrosymmetric superconductor li 2 pt 3 b},\ }\href@noop {} {\bibfield  {journal} {\bibinfo  {journal} {Physical Review B—Condensed Matter and Materials Physics}\ }\textbf {\bibinfo {volume} {86}},\ \bibinfo {pages} {134526} (\bibinfo {year} {2012})}\BibitemShut {NoStop}%
\bibitem [{\citenamefont {Xiao}\ \emph {et~al.}(2010)\citenamefont {Xiao}, \citenamefont {Chang},\ and\ \citenamefont {Niu}}]{xiao2010berry}%
  \BibitemOpen
  \bibfield  {author} {\bibinfo {author} {\bibfnamefont {D.}~\bibnamefont {Xiao}}, \bibinfo {author} {\bibfnamefont {M.-C.}\ \bibnamefont {Chang}},\ and\ \bibinfo {author} {\bibfnamefont {Q.}~\bibnamefont {Niu}},\ }\bibfield  {title} {\bibinfo {title} {Berry phase effects on electronic properties},\ }\href@noop {} {\bibfield  {journal} {\bibinfo  {journal} {Reviews of modern physics}\ }\textbf {\bibinfo {volume} {82}},\ \bibinfo {pages} {1959} (\bibinfo {year} {2010})}\BibitemShut {NoStop}%
\bibitem [{\citenamefont {Hagedorn}(1980)}]{hagedorn1980semiclassical}%
  \BibitemOpen
  \bibfield  {author} {\bibinfo {author} {\bibfnamefont {G.~A.}\ \bibnamefont {Hagedorn}},\ }\bibfield  {title} {\bibinfo {title} {Semiclassical quantum mechanics: I. the $\hbar\rightarrow 0$ limit for coherent states},\ }\href@noop {} {\bibfield  {journal} {\bibinfo  {journal} {Communications in Mathematical Physics}\ }\textbf {\bibinfo {volume} {71}},\ \bibinfo {pages} {77} (\bibinfo {year} {1980})}\BibitemShut {NoStop}%
\bibitem [{\citenamefont {Chang}\ and\ \citenamefont {Niu}(1996)}]{chang1996berry}%
  \BibitemOpen
  \bibfield  {author} {\bibinfo {author} {\bibfnamefont {M.-C.}\ \bibnamefont {Chang}}\ and\ \bibinfo {author} {\bibfnamefont {Q.}~\bibnamefont {Niu}},\ }\bibfield  {title} {\bibinfo {title} {Berry phase, hyperorbits, and the hofstadter spectrum: Semiclassical dynamics in magnetic bloch bands},\ }\href@noop {} {\bibfield  {journal} {\bibinfo  {journal} {Physical Review B}\ }\textbf {\bibinfo {volume} {53}},\ \bibinfo {pages} {7010} (\bibinfo {year} {1996})}\BibitemShut {NoStop}%
\bibitem [{\citenamefont {Son}\ and\ \citenamefont {Yamamoto}(2012)}]{son2012berry}%
  \BibitemOpen
  \bibfield  {author} {\bibinfo {author} {\bibfnamefont {D.~T.}\ \bibnamefont {Son}}\ and\ \bibinfo {author} {\bibfnamefont {N.}~\bibnamefont {Yamamoto}},\ }\bibfield  {title} {\bibinfo {title} {Berry curvature, triangle anomalies, and the chiral magnetic effect in fermi liquids},\ }\href@noop {} {\bibfield  {journal} {\bibinfo  {journal} {Physical review letters}\ }\textbf {\bibinfo {volume} {109}},\ \bibinfo {pages} {181602} (\bibinfo {year} {2012})}\BibitemShut {NoStop}%
\bibitem [{\citenamefont {Knoll}\ \emph {et~al.}(2020)\citenamefont {Knoll}, \citenamefont {Timm},\ and\ \citenamefont {Meng}}]{knoll2020negative}%
  \BibitemOpen
  \bibfield  {author} {\bibinfo {author} {\bibfnamefont {A.}~\bibnamefont {Knoll}}, \bibinfo {author} {\bibfnamefont {C.}~\bibnamefont {Timm}},\ and\ \bibinfo {author} {\bibfnamefont {T.}~\bibnamefont {Meng}},\ }\bibfield  {title} {\bibinfo {title} {Negative longitudinal magnetoconductance at weak fields in weyl semimetals},\ }\href@noop {} {\bibfield  {journal} {\bibinfo  {journal} {Physical Review B}\ }\textbf {\bibinfo {volume} {101}},\ \bibinfo {pages} {201402} (\bibinfo {year} {2020})}\BibitemShut {NoStop}%
\bibitem [{\citenamefont {Sharma}\ \emph {et~al.}(2023)\citenamefont {Sharma}, \citenamefont {Nandy}, \citenamefont {Raman},\ and\ \citenamefont {Tewari}}]{sharma2023decoupling}%
  \BibitemOpen
  \bibfield  {author} {\bibinfo {author} {\bibfnamefont {G.}~\bibnamefont {Sharma}}, \bibinfo {author} {\bibfnamefont {S.}~\bibnamefont {Nandy}}, \bibinfo {author} {\bibfnamefont {K.~V.}\ \bibnamefont {Raman}},\ and\ \bibinfo {author} {\bibfnamefont {S.}~\bibnamefont {Tewari}},\ }\bibfield  {title} {\bibinfo {title} {Decoupling intranode and internode scattering in weyl fermions},\ }\href@noop {} {\bibfield  {journal} {\bibinfo  {journal} {Physical Review B}\ }\textbf {\bibinfo {volume} {107}},\ \bibinfo {pages} {115161} (\bibinfo {year} {2023})}\BibitemShut {NoStop}%
\bibitem [{\citenamefont {Sondheimer}(1948)}]{sondheimer1948theory}%
  \BibitemOpen
  \bibfield  {author} {\bibinfo {author} {\bibfnamefont {E.}~\bibnamefont {Sondheimer}},\ }\bibfield  {title} {\bibinfo {title} {The theory of the galvanomagnetic and thermomagnetic effects in metals},\ }\href@noop {} {\bibfield  {journal} {\bibinfo  {journal} {Proceedings of the Royal Society of London. Series A. Mathematical and Physical Sciences}\ }\textbf {\bibinfo {volume} {193}},\ \bibinfo {pages} {484} (\bibinfo {year} {1948})}\BibitemShut {NoStop}%
\bibitem [{\citenamefont {Wang}\ \emph {et~al.}(2001)\citenamefont {Wang}, \citenamefont {Xu}, \citenamefont {Kakeshita}, \citenamefont {Uchida}, \citenamefont {Ono}, \citenamefont {Ando},\ and\ \citenamefont {Ong}}]{wang2001onset}%
  \BibitemOpen
  \bibfield  {author} {\bibinfo {author} {\bibfnamefont {Y.}~\bibnamefont {Wang}}, \bibinfo {author} {\bibfnamefont {Z.}~\bibnamefont {Xu}}, \bibinfo {author} {\bibfnamefont {T.}~\bibnamefont {Kakeshita}}, \bibinfo {author} {\bibfnamefont {S.}~\bibnamefont {Uchida}}, \bibinfo {author} {\bibfnamefont {S.}~\bibnamefont {Ono}}, \bibinfo {author} {\bibfnamefont {Y.}~\bibnamefont {Ando}},\ and\ \bibinfo {author} {\bibfnamefont {N.}~\bibnamefont {Ong}},\ }\bibfield  {title} {\bibinfo {title} {Onset of the vortexlike nernst signal above t c in la 2- x sr x cuo 4 and bi 2 sr 2- y la y cuo 6},\ }\href@noop {} {\bibfield  {journal} {\bibinfo  {journal} {Physical Review B}\ }\textbf {\bibinfo {volume} {64}},\ \bibinfo {pages} {224519} (\bibinfo {year} {2001})}\BibitemShut {NoStop}%
\bibitem [{\citenamefont {Behnia}(2009)}]{behnia2009nernst}%
  \BibitemOpen
  \bibfield  {author} {\bibinfo {author} {\bibfnamefont {K.}~\bibnamefont {Behnia}},\ }\bibfield  {title} {\bibinfo {title} {The nernst effect and the boundaries of the fermi liquid picture},\ }\href@noop {} {\bibfield  {journal} {\bibinfo  {journal} {Journal of Physics: Condensed Matter}\ }\textbf {\bibinfo {volume} {21}},\ \bibinfo {pages} {113101} (\bibinfo {year} {2009})}\BibitemShut {NoStop}%
\end{thebibliography}%
\end{document}